\documentclass[prd,preprint,tightenlines,floatfix,showpacs,preprintnumbers,nofootinbib,eqsecnum]{revtex4}

 \usepackage[dvips,final]{graphicx}
  \usepackage{amssymb}
   \usepackage{amsmath}
    \usepackage{amsfonts}
     \usepackage{epsfig}
      \usepackage{bm}

\newcommand{\GeV}{\textrm{GeV}}

\begin{document}

\begin{flushright}
LU TP 11-30\\
September 2011
\end{flushright}
\vfill
\title{New contributions to central exclusive production of dijets\\
in proton-(anti)proton collisions}

\author{Rafa{\l} Maciu{\l}a}
\email{rafal.maciula@ifj.edu.pl} \affiliation{Institute of Nuclear
Physics PAN, PL-31-342 Cracow, Poland}

\author{Roman Pasechnik}
 \email{Roman.Pasechnik@thep.lu.se}
 \affiliation{
Theoretical High Energy Physics, Department of Astronomy and
Theoretical Physics, Lund University, SE 223-62 Lund, Sweden}

\author{Antoni Szczurek}
\email{antoni.szczurek@ifj.edu.pl} \affiliation{Institute of Nuclear
Physics PAN, PL-31-342 Cracow,
Poland and\\
University of Rzesz\'ow, PL-35-959 Rzesz\'ow, Poland}

\date{\today}

\begin{abstract}
We consider central exclusive production of $gg$ dijets in
proton-proton (proton-antiproton) collisions at LHC and Tevatron for
different intermediate and final gluon polarisations. The amplitude
for the process is derived within the $k_\perp$-factorization
approach (with both the standard QCD and the Lipatov's effective
three-gluon verticies) and is considered in various kinematical
asymptotia, in particular, in the important limit of high-$p_\perp$
jets. Compared to earlier works we include emissions of gluons from
different gluonic $t$-channel lines as well as emission of
quark-antiquark dijets. Rapidity distributions, gluon jet $p_\perp$
distributions and invariant dijet mass distributions are presented.
We explore the competition of the standard diagram with both jets
emitted from a single $t$-channel gluon and the one with the
emission from both $t$-channel gluons. The second mechanism requires
a special treatment. We propose two different approaches. Including
special kinematics and using properties of off-diagonal gluons at
small $x$ and $\xi$ we arrive to correlations in two-dimensional
distributions in rapidity of one and second jet. We find that the
second contribution is much smaller than that known from the
literature. The digluon production constitutes an important
background to exclusive Higgs production.

\end{abstract}


\maketitle

\section{Introduction}

Experimental studies of hard exclusive processes, in particular, the
production of dijets at midrapidities, has been recently performed
at Tevatron \cite{CDF-dijets} and will be done at the LHC in the
near future \cite{CEP-review,CF09}.

It is expected that the contribution of the gluon pairs to the
exclusive hard dijets production dominates over that from
quark-antiquark pairs.
Martin, Ryskin and Khoze proposed a QCD mechanism of exclusive
digluon production \cite{MRK97}.
In certain regions of the phase space, the
process $pp\to p(gg)p$ (similarly to $q{\bar q}$ production) is
dominated by the non-perturbative region of gluon transverse
momenta, and even perturbative ingredients like the Sudakov form
factor are not under full theoretical control \cite{Cudell:2008gv}.
The problem becomes even more pronounced when considering the
irreducible backgrounds in central exclusive production (CEP) of Higgs
boson originating from the direct exclusive $b{\bar b}$ pair
production in a fusion of two off-shell gluons. In particular, in
Ref.~\cite{MPS-bbbar,MPS-higgs} it was shown that the central
exclusive production of $b{\bar b}$ jets at the LHC, may
noticeably shadow the corresponding signal of the Higgs boson in the
$b{\bar b}$ decay channel. Along with unknown NLO corrections to the
$k_{\perp}$-dependent hard subprocess amplitude $g^*g^*\to jj$ (in
particular, the NLO contribution from the rescattering of the final
state gluons into $q{\bar q}$ pairs can be potentially important),
this may lead to problems in experimental identification
of the Higgs boson.

Recently, it was shown in Ref.~\cite{DKRS2011} that the first LHC
measurements of the exclusive dijets would significantly reduce the
theoretical uncertainty for the central exclusive Higgs boson
production. This makes the process under consideration especially
important from both theoretical and experimental points of view.

Such a process has been recently investigated in detail in
Ref.~\cite{Cudell:2008gv}, and fairly good description of the
Tevatron data has been achieved. We would like to extend such an
analysis, both analytically and numerically, by analyzing separate
contributions from different final gluon polarisations, various
kinematical regions of the 4-particle phase space, which are
important for future LHC measurements, and the theoretical
uncertainties related with different choice of UGDFs and the
factorisation scale. Similarly to Ref.~\cite{Cudell:2008gv}, we
shall limit ourselves to the lowest-order QCD calculation, and
postpone the analysis of the higher order contributions for a
separate study.

Compared to the previous studies, we would like to perform an
estimation of the process when one gluon is emitted from a one
$t$-channel gluon of the QCD ladder and the second gluon is emitted
from the other $t$-channel gluon. This contribution was discussed in
Ref.~\cite{Cudell:2008gv} as potentially sizable. In this work, we
present the first numerical calculation of this contribution.

Recently, the calculation of the exclusive production of
quark-antiquark dijets has been performed for heavy $c \bar c$
\cite{MPS-ccbar} and $b \bar b$ \cite{MPS-bbbar,MPS-higgs} pairs.
Here, we extend our previous analysis and present the calculation
including both contributions from light ($u,\,d,\,s$) and heavy
($c,\,b$) quark/antiquark jets compared to the one from the gluonic
jets.

A high precision measurement of exclusive $b \bar b$ pair production
is required for central exclusive Higgs production measurements
\cite{CEP-review,CF09}. It is, therefore, instructive to estimate
the reducible background for Higgs CEP at LHC coming from
misidentification of gluon jets.

Another interesting point, which we would like to investigate here,
is the role of the gluon reggeization in the exclusive gluonic
dijets production. For this purpose, we employ the formalism of
the quasi-multi-Regge kinematics (QMRK) with the Lipatov's nonlocal
vertices for the triple-gluon coupling \cite{Lipatov}, and perform a
numerical comparison with the standard pQCD calculation  (with
standard gluons) \cite{Cudell:2008gv}.

This paper is organized as follows. In the second section, we
present the standard exclusive diffractive amplitude when both
gluons are produced from the same $t$-channel gluonic line as well
as the amplitude when one of the gluons is emitted from one line and
the second gluon from the second line. In the third section, we
briefly remind formulae for quark-antiquark dijets with arbitrary
quark mass. The fourth section contains discussion of unintegrated
gluon densities and model assumptions. In the Results section, we
present predictions for various differential distributions and compare our results with
the available CDF data as well as discuss corresponding theoretical uncertainties.
We also present predictions for future studies at LHC. Finally, the
summary and conclusions close our paper.

\section{Diffractive amplitude of the exclusive
gluon pair production}

\begin{figure}[h!]
\centerline{\epsfig{file=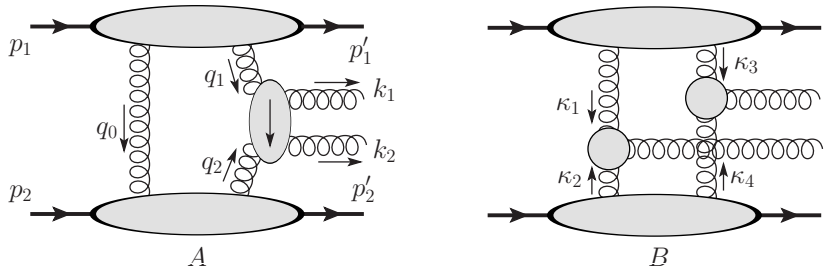,width=11.0cm}} \caption{Typical
diagrams for the exclusive gluon pair production in exclusive double
diffractive $pp$ scattering through the gluon-gluon fusion
subprocess $g^*g^*\to gg$ (A) and the 4-gluon fusion subprocess
$g^*g^*g^*g^*\to gg$ (B).} \label{fig:ggCEP}
\end{figure}

In this analysis, we apply the QCD mechanism for the central
exclusive production, proposed by the Durham group
(referred to below as the KMR approach) in Ref.~\cite{KMR_Higgs}. In
Fig.~\ref{fig:ggCEP} we show typical contributions to the exclusive
gluon pair production. In the important limit of high-$p_t$ jets,
the diagram B is suppressed by an extra hard propagator. Such a
limit has been considered in detail in Ref.~\cite{Cudell:2008gv}.
However, at relatively small gluon $p_\perp$'s, the diagram B may become sizeable and a
reliable numerical estimation of its contribution is required. In
this paper, for generality, we would like to calculate both
contributions in all potentially interesting regions of the
4-particle phase space.

Momenta of intermediate and final state gluons are given by the
following Sudakov decompositions in terms of incoming protons momenta
$p_{1,2}$
\begin{eqnarray}
&&q_1=x_1p_1+q_{1\perp},\quad q_2=x_2p_2+q_{2\perp},\quad
q_0=x'_1p_1+x'_2p_2+q_{0\perp}\simeq q_{0\perp},\quad x'_{1,2}\ll x_{1,2},\\
&&p_3=\beta_1p_1+\alpha_1p_2+k_{1\perp},\quad
p_4=\beta_2p_1+\alpha_2p_2+k_{2\perp}\,. \label{dec}
\end{eqnarray}
In forward scattering limit, we then have
\begin{eqnarray}
t_{1,2}=(p_{1,2}-p'_{1,2})^2={p'}^2_{1/2\perp}\to0,\qquad
q_{0\perp}\simeq-q_{1\perp}\simeq q_{2\perp}\,. \label{forward}
\end{eqnarray}
The Mandelstam invariants in the two-gluon fusion in the limit of
high-$p_{\perp}$ gluon jets $|{\bf k}|\equiv|{\bf p}_3|\simeq |{\bf
p}_4|\gg |{\bf q}_0|$ can be written as \cite{Cudell:2008gv}
\begin{eqnarray}
M_{gg}^2\equiv s_{gg}\simeq {\bf
k}^2\frac{(\beta_1+\beta_2)^2}{\beta_1\beta_2},\quad
t_{gg}\simeq-{\bf k}^2\frac{\beta_1+\beta_2}{\beta_1},\quad
u_{gg}\simeq-{\bf
k}^2\frac{\beta_1+\beta_2}{\beta_2}\,.\label{Mandel}
\end{eqnarray}
Here and below, we use notations for the transverse 2-momenta in
bold face style.

Let us first consider the explicit derivation of the diffractive
amplitude shown in Fig.~\ref{fig:ggCEP}(A) as an example. Starting
at the parton-level process and applying the cutting rules the
imaginary part of the one-loop partonic amplitude of the gluon pair
production with a fixed color indices $b_1$ and $b_2$ can be
calculated in the forward limit as (for similar derivation of Higgs
CEP amplitude, see Ref.~\cite{Forshaw05})
\begin{eqnarray}
&&\mathrm{Im}M_{b_1b_2}^{parton} = -\frac12\cdot 2\cdot
\frac{s}{2}\frac{1}{(2\pi)^2}\cdot\tau_{im}^{a}\tau_{jn}^{a}
\tau_{mk}^{c_1}\tau_{nl}^{c_2}f^{db_1c_1}f^{db_2c_2}\cdot
(2g_s)^4(p_1p_2)p_{1\rho}p_{2\sigma}\cdot(ig_s)^2\label{derivation}\\
&&\times\int\frac{d^2{\bf q}_0}{{\bf q}_0^2{\bf q}_1^2{\bf
q}_2^2}dx'_1dx'_2 \delta((p_1-q_0)^2)\delta((p_2+q_0)^2)\cdot{\cal
P}_1^{\rho\nu\beta}(q_1,r_1){\cal
P}_2^{\beta\mu\sigma}(r_1,-q_2)\cdot
\epsilon^*_{\mu}(\lambda_1)\epsilon^*_{\nu}(\lambda_2)\,, \nonumber
\end{eqnarray}
where the first factor $1/2$ comes from the cutting rule, factor 2
comes due to two identical contributing diagrams (emission of jets
from the first and second $t$-channel gluon line), factor
$\frac{s}{2}\frac{1}{(2\pi)^2}$ comes from the phase space in the
loop integration, ${\cal P}_{1,2}$ are the effective
Reggeon-Reggeon-gluon (RRG) verticies in the quasi-multi-Regge
kinematics (QMRK), corresponding to the kinematical configuration
with $\beta_1\gg \beta_2,\, \alpha_1\ll \alpha_2$ \cite{Lipatov}.
Here, we apply the eikonal approximation for the quark-gluon
verticies in the proton defined by
$2g_s\tau^a_{ij}p_{1,2}\delta_{\lambda\lambda'}$, where $\tau^a$ are
the Gell-Mann matricies, $g_s$ is the QCD coupling, and
$\delta_{\lambda\lambda'}$ appears due to the fact that soft gluons
cannot change quark helicity in an energetic proton.

For the color singlet production (since in- and outgoing protons are
in the color singlet state) there is no color transfer
from quark lines between protons, i.e. color indices in initial and
final quarks are the same, i.e. $i=k$ and $j=l$. Then, the color
averaging in each quark line leads to the substitution
\begin{eqnarray}\nonumber
\tau_{im}^{a}\tau_{jn}^{a}
\tau_{mk}^{c_1}\tau_{nl}^{c_2}\quad\to\quad\frac{\delta^{c_1c_2}}{4N_c^2}
\end{eqnarray}
in the diffractive amplitude (\ref{derivation}). The appearance of
$\delta^{c_1c_2}$ here automatically gives rise to the projection of
the produced $gg$-pair onto the color singlet state.

Further, in order to go over to the hadron level, one has to absorb
the factor
\begin{eqnarray}
\frac{C_F\alpha_s}{\pi}=\frac{N_c^2-1}{2N_c}\frac{\alpha_s}{\pi},\quad
\alpha_s=\frac{g_s^2}{4\pi}
\end{eqnarray}
into a definition of the unintegrated ($q_{\perp}$-dependent) gluon
distribution function (UGDF) along each proton line as required by
the underlying $k_\perp$-factorisation approach \cite{Forshaw05}. A
reliable model for generalized off-diagonal UGDFs used in the current
analysis will be discussed in some detail below.

Due to the gauge invariance and the factorisation property of the
hard RRG vertices \cite{Lipatov} we get
\begin{eqnarray}\nonumber
p_{1\rho}p_{2\sigma}{\cal P}_1^{\rho\nu\beta}(...){\cal
P}_2^{\beta\mu\sigma}(...)=\frac{s}{4}n^+_{\rho}n^-_{\sigma}{\cal
P}_1^{\rho\nu\beta}(...){\cal
P}_2^{\beta\mu\sigma}(...)\quad\to\quad \frac{s}{4}
\frac{C_1^{\nu}(...)\,C_2^{\mu}(...)}{{\bf r}_1^2},\quad r_1=q_1-p_3
\; ,
\end{eqnarray}
where $n^{\pm}=p_{1,2}/E_p^{cms}$, $E_p^{cms}=\sqrt{s}/2$, and
$C_{1,2}^{\mu}$ are the nonlocal RRG couplings defined as
\cite{Lipatov}
\begin{eqnarray}\nonumber
&&C_1^{\mu}(v_1,v_2)=p_1^{\mu}\left(\beta_1-\frac{2{\bf
v}_1^2}{s\alpha_1}\right)-p_2^{\mu}\left(\alpha_1-\frac{2{\bf
v}_2^2}{s\beta_1}\right)-(v_{1\perp}+v_{2\perp})^{\mu},\\
&&C_2^{\mu}(v_1,v_2)=p_1^{\mu}\left(\beta_2-\frac{2{\bf
v}_1^2}{s\alpha_2}\right)-p_2^{\mu}\left(\alpha_2-\frac{2{\bf
v}_2^2}{s\beta_2}\right)-(v_{1\perp}+v_{2\perp})^{\mu}\,.
\label{Lipatov}
\end{eqnarray}

Finally, the contributions of diagrams Fig.~\ref{fig:ggCEP} (A) and
(B) to the diffractive amplitude ${\cal M}^{gg} = {\cal M}^{A} +
{\cal M}^{B}$ for the central exclusive $gg$ (with external color
indices $a$ and $b$) dijet production $pp\to p(gg)p$ read
\begin{eqnarray}\nonumber
{\cal M}^{A}_{ab}(\lambda_1,\lambda_2)&=&is\,{\cal
A}\,\frac{\delta_{ab}}{N_c^2-1}\int d^2{\bf
q}_0\frac{f^{\mathrm{off}}_g(q_0,q_1)f^{\mathrm{off}}_g(q_0,q_2)\cdot
\epsilon^*_{\mu}(\lambda_1)\epsilon^*_{\nu}(\lambda_2)}{{\bf
q}_0^2{\bf q}_1^2{\bf q}_2^2}\,\times\\
&&\left[\frac{C_1^{\mu}(q_1,r_1)C_2^{\nu}(r_1,-q_2)}{{\bf r}_1^2}
+\frac{C_1^{\mu}(q_1,r_2)C_2^{\nu}(r_2,-q_2)}{{\bf r}_2^2}\right]\,,\label{ampl-a+b}\\
{\cal M}^{B}_{ab}(\lambda_1,\lambda_2)&=&-is\,{\cal
A}\,\frac{\delta_{ab}}{N_c^2-1}\int d^2{\bm
\kappa}_1\frac{f^{\mathrm{off}}_g(\kappa_1,\kappa_3)f^{\mathrm{off}}_g(\kappa_2,\kappa_4)\cdot
\epsilon^*_{\mu}(\lambda_1)\epsilon^*_{\nu}(\lambda_2)}{{\bm
\kappa}_1^2{\bm \kappa}_2^2{\bm \kappa}_3^2{\bm \kappa}_4^2}\,\times\nonumber \\
&&C_1^{\mu}(\kappa_1,-\kappa_2)C_2^{\nu}(\kappa_3,-\kappa_4),
\label{ampl-c}
\end{eqnarray}
where ${\cal A}=2\pi^2g_s^2/C_F$, the minus sign in ${\cal M}^{B}$
comes from the difference in colour factors,
$f^{\mathrm{off}}_g(v_1,v_2)$ is the off-diagonal UGDF, which is
dependent on longitudinal and transverse components of both gluons
with 4-momenta $v_1$ and $v_2$, emitted from a single proton line,
and
\[r_2=q_1-p_4\,,\quad\kappa_2=-(\kappa_1-p_4)\,,\quad\kappa_4=-(\kappa_3-p_3)\,.\]
Then the matrix element squared for the exclusive diffractive $gg$
production cross section can be written in the standard way
\begin{eqnarray}
|{\cal M}|^2=\sum_{a,b}{\cal M}_{ab}({\cal M}_{ab})^*\,,
\end{eqnarray}
summing up over all possible color singlet combinations of the final
gluons.

The integration in the case of the emission from different
$t$-channel gluon lines (diagram B in Fig.~\ref{fig:ggCEP}) can be
made symmetric with respect to both protons by the following
equivalent transformation of the integral measure
\begin{equation}
\int d^2 {\bm \kappa}_1 \quad \to \quad \frac{1}{4} \int d^2 {\bm
\kappa}_{-} d^2 {\bm \kappa}_{+} \delta^{2}\left( {\bm \kappa}_{+} -
{\bf p}_3 \right)\,, \label{symmetric_integration}
\end{equation}
where ${\bm \kappa}_- = {\bm \kappa}_1 - {\bm \kappa}_2$ and ${\bm
\kappa}_+ = {\bm \kappa}_1 + {\bm \kappa}_2$ have been introduced.
In practice, such a transformation is convenient in numerical
calculations below.

When the $p_\perp$'s of the final jets are sufficiently large, the
contribution of the diagram B should vanish much faster than that
of the diagram A due to an extra propagator suppression in the
amplitude (see Eq.~\ref{ampl-c}). Moreover, the hard transverse
momentum flow through the proton remnant would disturb it too much
such that it becomes less likely to combine it back to an exclusive
proton-like state after hadronisation, which should be reflected in
an extra suppression by the UGDFs behavior at large gluon $p_\perp$. Such
arguments lead to a conclusion that the diagram B can be
sizeable only at relatively small jet $p_\perp$'s, but the large invariant
mass of the $gg$ dijet system (i.e. at large rapidity difference
$\Delta y=|y_1-y_2|$ between two jets at the edges of the central
detector). Numerical estimation of such a contribution could be,
therefore, important when the statistics on diffractive dijets
production at LHC becomes sufficiently large. Potentially, it could
even be singled out and its features could be tested if there is a
region in the phase space where it may dominate.

For the emission of both gluons from the same $t$-channel gluon line
(the standard CEP process given by diagram A in
Fig.~\ref{fig:ggCEP}) we have typically: $q_{1\perp} \sim q_{2\perp}
\ll p_{3\perp} \sim p_{4\perp}$. The integration over screening
gluon transverse momentum is limited to rather small
$q_{0\perp}\simeq q_{1/2\perp}$ (in the forward limit).

The kinematical situation for the diagram B is different. Typically,
in this case either $\kappa_{1\perp}$ is large and of the order of
$p_{3\perp}$ and $\kappa_{2\perp}$ is small, or vice versa ---
$\kappa_{1\perp}$ is small and $\kappa_{2\perp}$ is large and of the
order of $p_{3\perp}$. The integration over $\kappa_-$ extends to
large values, which means that typical transverse momenta of gluons
in the impact factors are large. So, technically, when using grids
for UGDFs we have to do it separately for both situations. Also, the
kinematical structure of UGDFs are very different in diagrams A and
B. This issue will be discussed in detail below.

In Eqs.~(\ref{ampl-a+b}) and (\ref{ampl-c}),
$\epsilon^*_{\mu}(\lambda_1)$ and $\epsilon^*_{\nu}(\lambda_2)$ are
the polarisation vectors of the final state gluons with helicities
$\lambda_1,\,\lambda_2$ and momenta $p_3,\,p_4$, respectively. They
can be defined in the $gg$ rest frame with $z$ axis along the proton
beam as
\begin{eqnarray}\nonumber
&&\epsilon^*_{\mu}(\lambda_1)=-\frac{1}{\sqrt{2}}
(0,\,\lambda_1\cos\theta\cos\psi-i\sin\psi,\,
\lambda_1\cos\theta\sin\psi+i\cos\psi,\,-\lambda_1\sin\theta)\,,\\
&&\epsilon^*_{\nu}(\lambda_2)=-\frac{1}{\sqrt{2}}
(0,\,-\lambda_2\cos\theta\cos\psi-i\sin\psi,\,
-\lambda_2\cos\theta\sin\psi+i\cos\psi,\,\lambda_2\sin\theta)\,,\label{glupol}
\end{eqnarray}
such that
$\epsilon^{\mu}(\lambda_1)\epsilon^*_{\mu}(\lambda_2)=-\delta^{\lambda_1,-\lambda_2}$
and
$\epsilon^*_{\mu}(\lambda_1)p_3^{\mu}=\epsilon^*_{\nu}(\lambda_2)p_4^{\nu}=0$.
In this frame, momenta of protons and final-state gluons are
\begin{eqnarray}\nonumber
&&p_1^{\mu}=\frac{E_1}{\sqrt{2}}(1,\,0,\,0,\,1),\qquad
p_2^{\mu}=\frac{E_2}{\sqrt{2}}(1,\,0,\,0,\,-1),\\
&&p_3^{\mu}=E_g(1,\,\sin\theta\cos\psi,\,\sin\theta\sin\psi,\,\cos\theta),\label{frame}\\
&&p_4^{\nu}=E_g(1,\,-\sin\theta\cos\psi,\,-\sin\theta\sin\psi,\,-\cos\theta),\nonumber
\end{eqnarray}
so that the proton and gluon energies $E_{1,2},\,E_g$ and the polar
angle of a gluon jet $\theta$ w.r.t. the $z$-axis are defined as
\begin{eqnarray*}
E_g\equiv\frac{\sqrt{s_{gg}}}{2}=\frac{E_1}{\sqrt{2}}(\beta_1+\beta_2)=
\frac{E_2}{\sqrt{2}}(\alpha_1+\alpha_2),\quad
\cos\theta=\frac{\beta_1-\beta_2}{\beta_1+\beta_2},\quad
\sin\theta=\frac{2\sqrt{\beta_1\beta_2}}{\beta_1+\beta_2}\,.
\end{eqnarray*}

In the high-$p_{\perp}$ limit and at central rapidities of jets, the
$gg$ dijet rest frame, introduced above, becomes identical to the
initial protons c.m.s. frame, which we use in actual numerical
calculations below. Then the diffractive amplitude (\ref{ampl-a+b})
reduces to the standard expression with $gg\to gg$ hard scattering
amplitude initially derived in Ref.~\cite{Cudell:2008gv}
\begin{eqnarray}\nonumber
{\cal M}^{A}_{ab}&\simeq& 2i{\cal A}\frac{s}{{\bf
k}^2}\frac{\delta_{ab}}{N_c^2-1}\int d^2{\bf
q}_0\frac{f^{\mathrm{off}}_g(q_0,q_1)f^{\mathrm{off}}_g(q_0,q_2)}{{\bf
q}_0^2{\bf q}_1^2{\bf
q}_2^2}\sum_{\lambda_1^*\lambda_2^*}e^{i(\lambda_1^*-\lambda_2^*)\phi}
\times\\&&(-2\lambda_1^*\lambda_2^*)|{\bf q}_1||{\bf
q}_2|e^{-i\lambda_1^*\phi_1+i\lambda_2^*\phi_2}\cdot
A(\lambda_1^*\lambda_2^*\to \lambda_1\lambda_2) \; ,
\label{amp-Cudell}
\end{eqnarray}
where $\phi_{1,2}$ and $\phi$ are the azimuthal angles of the fusing
gluons $q_{1,2}$ and of the dijet production plane, respectively,
$\lambda_1^*,\,\lambda_2^*$ are the helicities of fusing gluons with
momenta $q_1,\,q_2$, respectively, and the nonzeroth helicity
amplitudes
\begin{eqnarray*}
&&A(++\to++)=A(--\to--)=1 \; ,\\
&&A(+-\to+-)=A(-+\to-+)=\frac{u_{gg}^2}{s_{gg}^2} \;,\\
&&A(+-\to-+)=A(-+\to+-)=\frac{t_{gg}^2}{s_{gg}^2}\,,
\end{eqnarray*}
with Mandelstam invariants defined in Eq.~(\ref{Mandel}), from which
we see that in the QMRK limit amplitudes
$A(\lambda_1^*\lambda_2^*\to \lambda_1\lambda_2)$ reduce simply to
$\delta_{\lambda_1^*\lambda_1}\delta_{\lambda_2^*\lambda_2}$. In the
forward limit, provided by Eq.~(\ref{forward}), we have
$\phi_2\simeq \phi_1+\pi$, so the integral in Eq.~(\ref{amp-Cudell})
\begin{eqnarray*}
\int d^2{\bf q}_0\,e^{-i(\lambda_1-\lambda_2)\phi_1}
\end{eqnarray*}
survives only when $\lambda_1=\lambda_2$, i.e. when the $gg$ dijet
is produced in the $J_z=0$ state, which corresponds to the
well-known $J_z=0$ selection rule in the central exclusive
production processes \cite{Khoze:2000jm}. We will discuss the
subleading corrections to this rule below when presenting the
numerical results for different gluon polarizations.

Below, for consistency, in order to see effects of the gluon
reggeization and subleading corrections to the high-$p_{\perp}$
limit we numerically compare both versions of the diffractive
amplitudes -- in the QMRK approximation (\ref{ampl-a+b}) and in the
standard approach (\ref{amp-Cudell}), as well as estimate subleading
contribution to the observable signal from the amplitude B with
symmetric gluon couplings (\ref{ampl-c}). The bare amplitudes above
are subjected to absorption corrections which depend on collision
energy and typical proton transverse momenta. We shall discuss this
issue shortly when presenting our results.

\section{Quark-antiquark dijets production}

Let us consider now the contribution of the quark/antiquark pairs to
the observable signal of the exclusive dijets production. The hard
subprocess amplitude for the $q\bar{q}$ pair production via
off-shell gluon-gluon fusion was previously discussed in detail in
Refs.~\cite{MPS-bbbar,MPS-higgs}. Here we would like to list the
relevant formulae only, which will be used in numerical calculations
below.

The amplitude of the exclusive diffractive $q\bar{q}$ pair
production $pp\to p(q\bar{q})p$ reads \cite{MPS-higgs}
\begin{eqnarray}
{\cal M}_{q{\bar q}}(\lambda_1,\lambda_2)&=&i\,s\cdot4\pi^2
\frac{\delta_{c_1c_2}}{(N_c^2-1)^2}\,\int d^2{\bf
q}_0\,V_{\lambda_q\lambda_{\bar{q}}}^{c_1c_2}\,
\frac{f^{\mathrm{off}}_g(q_0,q_1)f^{\mathrm{off}}_g(q_0,q_2)}{{\bf
q}_0^2{\bf q}_1^2{\bf q}_2^2}\,, \label{ampl}
\end{eqnarray}
Here, the vertex factor $V_{\lambda_q\lambda_{\bar{q}}}^{c_1c_2}$ is
the production amplitude of a pair of massive quark $q$ and
antiquark $\bar{q}$ with helicities $\lambda_q$, $\lambda_{\bar{q}}$
and momenta $p_3$, $p_4$, respectively. It is given by the following
general expression
\begin{eqnarray}\nonumber
&&{}V_{\lambda_q\lambda_{\bar{q}}}^{c_1c_2}=-\frac{2g^2}{M_{q\bar{q}\perp}^2\sqrt{N_c}}\,
\delta^{c_1c_2}\,\bar{u}_{\lambda_q}(p_3)
\biggl(\frac{\hat{q}_{1\perp}\hat{q}_{1}-2(p_{3\perp}q_{1\perp})}
{q_{1\perp}^2-2(p_3q_1)}\,\hat{q}_{2\perp}-\hat{q}_{2\perp}\,\frac{\hat{q}_{1}\hat{q}_{1\perp}-
2(p_{4\perp}q_{1\perp})}{q_{1\perp}^2-2(p_4q_1)}\biggr)v_{\lambda_{\bar{q}}}(p_4).\\
\label{fin-vert-fin}
\end{eqnarray}

In analogy with Eq.~(\ref{dec}), one can introduce the Sudakov
expansions for quark momenta as
\begin{eqnarray}
p_3=x_1^qp_1+x_2^qp_2+p_{3\perp},\quad p_4=x_1^{\bar q}p_1+x_2^{\bar
q}p_2+p_{4\perp}
\end{eqnarray}
leading to
\begin{eqnarray}\label{xqq}
x_{1,2}=x_{1,2}^q+x_{1,2}^{\bar q},\quad
x_{1,2}^q=\frac{m_{1\perp}}{\sqrt{s}}e^{\pm y_1},\quad x_{1,2}^{\bar
q}=\frac{m_{2\perp}}{\sqrt{s}}e^{\pm y_2},\quad
m_{1/2\perp}^2=m_q^2+|{\bf k}_{1/2\perp}|^2\,,
\end{eqnarray}
in terms of quark/antiquark rapidities $y_1$, $y_2$ and transverse
masses $m_{1\perp}$, $m_{2\perp}$. The only difference of the
quark/antiquark fractions $x_{1,2}^{q,\bar{q}}$ from that of the
gluons is that they are dependent on the quark mass $m_q$.

It is convenient to fix the c.m.s. frame of the $q\bar{q}$ pair with
$z$ axis along the proton beam, so ${\bf p}_3=-{\bf p}_4={\bf k}$
and $p_{3,4}^0=M_{q{\bar q}}/2$. The gluon and quark transverse
momenta (with respect to the proton beam) in the polar coordinates
are then defined as
\begin{eqnarray}\nonumber
{\bf q}_{0\perp}=q_{\perp}(\cos\psi,\sin\psi),\qquad {\bf
p}_{3\perp}=-{\bf
p}_{4\perp}=k_{\perp}(\cos\varkappa,\sin\varkappa),\quad
\end{eqnarray}
respectively, and
\begin{eqnarray}\label{inv-kin}
k_{\perp}=E_q\frac{\sqrt{\gamma^2(x_1^q+x_1^{\bar
q})^2-(x_1^q-x_1^{\bar q})^2}}{x_1^q+x_1^{\bar q}},\quad
k_z=E_q\frac{x_1^q-x_1^{\bar q}}{x_1^q+x_1^{\bar q}},\quad |{\bf
k}|=\sqrt{k_{\perp}^2+k_z^2}=E_q\gamma\,.
\end{eqnarray}

In these notations, the helicity amplitudes $g^*g^*\to
q(\lambda_1)\bar{q}(\lambda_2)$ can be written as (for more detail,
see Ref.~\cite{MPS-higgs})
\begin{eqnarray}\nonumber
&&V_{+-}={\cal C}\,\frac{q_{\perp}^2}{|{\bf k}|}\,\bigg[2|{\bf
k}|q_{\perp}\Big(|{\bf
k}|\cos(\psi-\varkappa)-ik_z\sin(\psi-\varkappa)\Big)+
M_{q\bar{q}}k_{\perp}\Big(k_z\cos(2\psi-2\varkappa)-\\&&i|{\bf
k}|\sin(2\psi-2\varkappa)\Big)\bigg]/
\bigg[M_{q\bar{q}}^2(k_{\perp}^2+q_{\perp}^2+m_q^2)+4M_{q\bar{q}}k_{\perp}q_{\perp}k_z\cos(\psi-\varkappa)-\nonumber\\&&
2k_{\perp}^2q_{\perp}^2(1+\cos(2\psi-2\varkappa))+q_{\perp}^4\bigg],\label{Vpm}\\
&&V_{++}=-2{\cal C}\,e^{-i\varkappa}\frac{q_{\perp}^2m_q}{|{\bf
k}|}\,\bigg[k_{\perp}^2\cos(2\psi-2\varkappa)+|{\bf k}|^2\bigg]/
\bigg[M_{q\bar{q}}^2(k_{\perp}^2+q_{\perp}^2+m_q^2)+\nonumber\\&&4M_{q\bar{q}}k_{\perp}q_{\perp}k_z\cos(\psi-\varkappa)-
2k_{\perp}^2q_{\perp}^2(1+\cos(2\psi-2\varkappa))+q_{\perp}^4\bigg]\label{Vpp}
\end{eqnarray}
where $m_q$ is the quark mass, and the normalisation factor is
${\cal C}=2g^2\delta^{c_1c_2}/\sqrt{N_c}$. Below, we will use the
expressions (\ref{Vpm}) and (\ref{Vpp}) in calculations of the quark
jets contribution to the central exclusive dijets production at
Tevatron and LHC energies of both light ($u,\,d,\,s$) and heavy
($c,\,b$) quarks.

\section{Off-diagonal unintegrated gluon distributions}

\subsection{Emission from the same $t$-channel gluon}

The off-diagonal unintegrated gluon distribution in
Eq.~(\ref{ampl-a+b}), where longitudinal momentum fractions satisfy
the strong inequality $x' \ll x$, is calculated in the forward
scattering limit $q_{0\perp}\simeq q_{1/2\perp}$ according to the
Kimber-Martin-Ryskin (KMR) prescription \cite{KMR2001,MR}
\begin{equation}
f^{\mathrm{off}}_{1/2g}(x_{1,2},x',q_{1/2\perp}^2,q_{0\perp}^2,\mu^2;t)
= R_g(x') \frac{d}{d\ln q_{\perp}^2} \left( x_{1,2}
g(x_{1,2},q_{\perp}^2) S(q_{\perp}^2,\mu^2) \right) |_{q_{\perp}^2 =
q_{1/2\perp}^2} \cdot F(t)\,, \label{off-diagonal-UGDF}
\end{equation}
which was used e.g. in exclusive dijets studies in
Refs.~\cite{DKRS2011} and leads to a reasonable description of the
Tevatron data \cite{CDF-dijets}. In the equation above, $x_{1,2}$
and $q_{1/2\perp}^2$ are longitudinal momentum fractions with
respect to the parent proton and transverse momenta squared of the
active gluons $q_{1\perp}$ and $q_{2\perp}$, respectively, $x'\sim
q_{0\perp}/\sqrt{s}$ and $q_{0\perp}^2$ are the same variables for
the screening gluon $q_{0\perp}$.
If we assume that at small $x$: $x g(x) = N_g x^{\lambda_g}$
the skewedness parameter can be expressed in terms of the $\lambda_g$
as \cite{Shuvaev:1999ce}:
\begin{equation}
R_g = \frac{2^{2 \lambda_g+3}}{\sqrt{\pi}}
\frac{\Gamma(\lambda_g+5/2)}{\Gamma(\lambda_g+4)} \; .
\label{R_g}
\end{equation}
We will take $R_g\simeq 1.2$ in practical calculations.
The function $xg(x,q_{\perp}^2)$ in Eq.~(\ref{off-diagonal-UGDF}) is the
collinear DGLAP gluon distribution, $S(q_{\perp}^2,\mu^2)$ is the
so-called Sudakov form factor and the nucleon form factor in the
forward limit $F(t)=\exp(bt/2)$ with the slope parameter $b\simeq
4\,\GeV^{-2}$ \cite{KMR00} describes the coupling of the gluonic
ladders to one of the nucleon lines (see Fig.~\ref{fig:ggCEP}).

The range of the integration over $d^2 {\bf q}_0$ is formally
limited by the existence of the DGLAP gluon PDF $xg(x,q_{\perp}^2)$.
In explicit calculations we use the next-to-leading order CTEQ6
collinear distributions \cite{CTEQ6} for which $xg(x,\mu^2)$
parametrization works well down to quite small $\mu_0^2 \sim$ 0.4
GeV$^2$. This way we cut off the region below this starting scale
$\mu_0^2$. This has not practical consequences for production of
relatively large invariant masses as these regions are also
numerically suppressed by the behavior of the Sudakov form factor
for small $q_{\perp}^2$. In order to exhibite uncertainties of our numerical results
related to the collinear PDFs we also use GRV94 \cite{GRV94}, GJR08 \cite{GJR08}
and MSTW08 \cite{MSTW08} distributions.

The longitudinal momentum fractions of the fusing gluons entering
Eq.~(\ref{off-diagonal-UGDF}) are calculated as
\begin{eqnarray}
x_1 &=& \frac{p_{3\perp}}{\sqrt{s}} \exp(+y_3)
     +  \frac{p_{4\perp}}{\sqrt{s}} \exp(+y_4) \; , \nonumber \\
x_2 &=& \frac{p_{3\perp}}{\sqrt{s}} \exp(-y_3)
     +  \frac{p_{4\perp}}{\sqrt{s}} \exp(-y_4) \; 
\label{fractional_momentum_fractions_diagrama}
\end{eqnarray}
from the transverse momenta $p_{3/4,\perp}$ and rapidities $y_{3,4}$
of the final gluonic jets.

\subsection{Emission from both $t$-channel gluons}

The kinematics of the diagram B in Fig.~\ref{fig:ggCEP} is different
from that for the diagram A. Here, the off-diagonal unintegrated
gluon distributions $f^{\mathrm{off}}_g(x_1,x_3,{\bm
\kappa}_1^2,{\bm \kappa}_3^2,\mu_1^2,\mu_2^2)$ and
$f^{\mathrm{off}}_g(x_2,x_4,{\bm \kappa}_2^2,{\bm
\kappa}_4^2,\mu_1^2,\mu_2^2)$ should be evaluated at $x_1 \sim x_2$
and $x_3 \sim x_4$. In general, such objects are not well known and
were not discussed so far in the literature. We calculate the
longitudinal momentum fractions of the fusing gluons in the
considered kinematical domain as follows
\begin{eqnarray}
x_1 \simeq \frac{p_{3\perp}}{\sqrt{s}} \exp(+y_3) \, , \qquad x_2
\simeq \frac{p_{4\perp}}{\sqrt{s}} \exp(-y_3)
\, , \nonumber \\
x_3 \simeq \frac{p_{3\perp}}{\sqrt{s}} \exp(+y_4) \, , \qquad x_4
\simeq \frac{p_{4\perp}}{\sqrt{s}} \exp(-y_4) \, .
\label{longitudinal_momentum_fractions_diagramb}
\end{eqnarray}
As a first approximation, one could try to use a symmetric
factorized prescription for the off-diagonal UGDFs, which was
successfully used before for the exclusive production of $\chi_c$
mesons in Ref.~\cite{PST_chic}
\begin{eqnarray}
f^{\mathrm{off}}_g(x_1,x_3,{\bm \kappa}_1^2,{\bm
\kappa}_3^2,\mu_1^2,\mu_2^2;t) &=& \sqrt{ f_g(x_1,{\bm
\kappa}_1^2,\mu_1^2) f_g(x_3,{\bm \kappa}_3^2,\mu_2^2) } \cdot
F(t_1)
\; , \nonumber \\
f^{\mathrm{off}}_g(x_2,x_4,{\bm \kappa}_2^2,{\bm
\kappa}_4^2,\mu_1^2,\mu_2^2;t) &=& \sqrt{ f_g(x_2,{\bm
\kappa}_2^2,\mu_1^2) f_g(x_4,{\bm \kappa}_4^2,\mu_2^2) } \cdot
F(t_2) \; . \label{off-diagonal-UGDFs-diagramb}
\end{eqnarray}
Above unintegrated diagonal distributions include Sudakov form
factors in the same way as in the KMR UGDF
(\ref{off-diagonal-UGDF}). Since for the jet production in the
diagram B $p_{3\perp}
> \kappa_{1\perp},\,\kappa_{2\perp}$ and $p_{4\perp} >
\kappa_{3\perp},\,\kappa_{4\perp}$ in most cases, so a physically
reasonable choice of scales in the scale-dependent UGDFs would be
$\mu_1 = p_{3\perp}$ and $\mu_2 = p_{4\perp}$ or $\mu_1 = \mu_2 =
M_{jj}$. We adopt these simplest choices since we do not know the
exact evolution of the Sudakov form factor in the considered
kinematical domain (see, also a discussion of this issue in
Ref.~\cite{Cudell:2008gv}). In this case, when the gluon $q_{\perp}$
becomes bigger than the scale $\mu$, we take simply $S(q_{\perp}^2,
\mu^2) = 1$.

Our prescription in Eq.~(\ref{off-diagonal-UGDFs-diagramb}) does not
use the fact that in the considered process both gluons are
outgoing (emitted). In the collinear approach this corresponds to the ERBL
kinematical region \cite{ERBL} where
\[ |x| < |\xi|,\qquad x = \frac{x_1 + x_2}{2},\qquad \xi = \frac{x_1 -
x_2}{2}\,. \]
In our case of central dijet production typically both $x_1$ and
$x_2$ are small that is also $x$ and $\xi$ are small. In this region
the collinear off-diagonal distributions $H(x,\xi,\mu^2,t)$ can be
estimated in a model independent way \cite{Shuvaev:1999ce}.

The discussion above suggests therefore another prescription for UGDFs in this special kinematical case:
\begin{equation}
f^{\mathrm{off}}_g(x,x',k_t^2,{k'}_t^2,\mu^2,t) =
R_{coll}(x,x';\mu^2,t=0)\cdot\sqrt{f_g(\bar{x},k_t^2,\mu^2)
f_g(\bar{x},{k'}_t^2,\mu^2) } \cdot F(t) \; ,
\label{new_prescription}
\end{equation}
where $\bar{x} = \frac{x + x'}{2}$, $\mu^2=\mu_1^2\simeq \mu_2^2$
and $f_g$ are standard diagonal unintegrated distributions as in e.g.
Ref.~\cite{KMR2001,MR}. Here $R_{coll}$ is the ratio of collinear off-diagonal
distributions in ERBL to DGLAP region:
\begin{equation}
R_{coll}(x_1,x_2;\mu^2,t=0) =
\frac{H^{ERBL}_g(x,\xi;\mu^2,t=0)}{H^{DGLAP}_g(x,\xi;\mu^2,t=0)} \; .
\label{ratio}
\end{equation}

Assuming that at small $x$: $x g(x) = N_g x ^{-\lambda_g}$
in the limit of small $x$ and $\xi$ the off-diagonal distribution $H_g(x,\xi,t)$ can be expressed
in terms of $\lambda_g$ as:
\begin{equation}
H_g(x,\xi,t) = N_g \frac{\Gamma(\lambda_g + 5/2)}{\Gamma(\lambda_g+2)}
\frac{2}{\sqrt{\pi}}
\int_{0}^{1} ds [x+\xi(1-2s)]
\left( \frac{4 s (1-s) }{ x + \xi(1-2s)} \right)^{\lambda_g+1}   \; .
\label{small_xxi_H_g}
\end{equation}
For our estimates here $\lambda_g$ is a crucial parameter which is not completely well known.
In the double logarithm approximation at small values of $x$:
\begin{equation}
\lambda_g = \sqrt{\frac{\alpha_s(\mu^2)}{\pi}
\log\left(\frac{1}{x}\right) \log\left(\frac{\mu^2}{\mu_0^2}\right) }  \; .
\label{lambda_g_double_log}
\end{equation}
However, the gluon distribution at $x <$ 10$^{-4}$ and small factorization scales
is poorly known (see e.g. a discussion in \cite{MPS-bbbar}). In consequence
applicability of the double logarithmic formula (\ref{lambda_g_double_log}) is not obvious and not well
justified.
Therefore we will treat $\lambda_g$ as a free parameter. In general, it can be dependent
on the scale of the problem (transverse momentum of the jet). To demonstrate uncertainties we shall show
results for $\lambda_g$ = 0.2, 0.4, 0.6 and 0.8.

\begin{figure}[!h]
 \includegraphics[width=6.0cm]{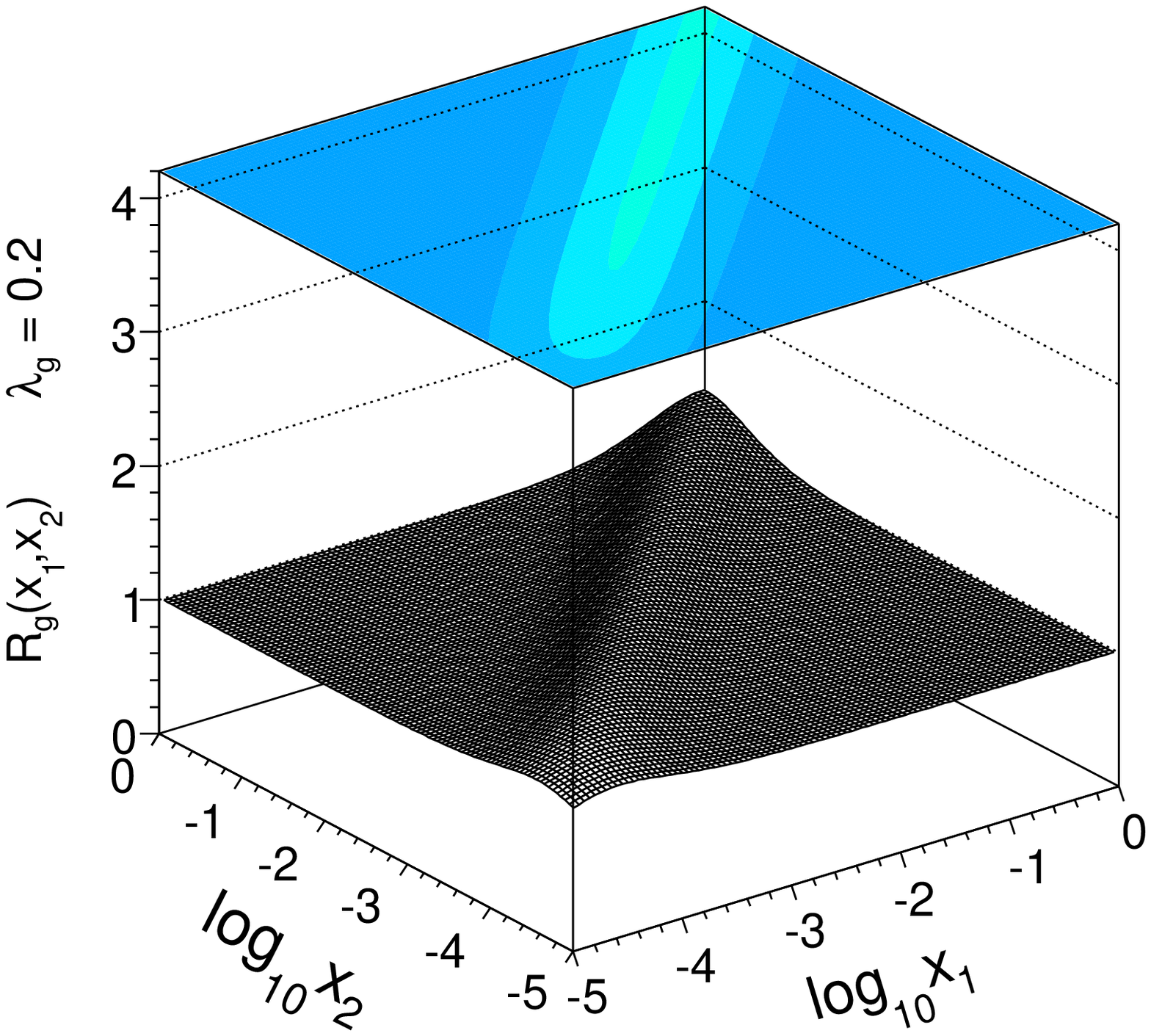}
 \includegraphics[width=6.0cm]{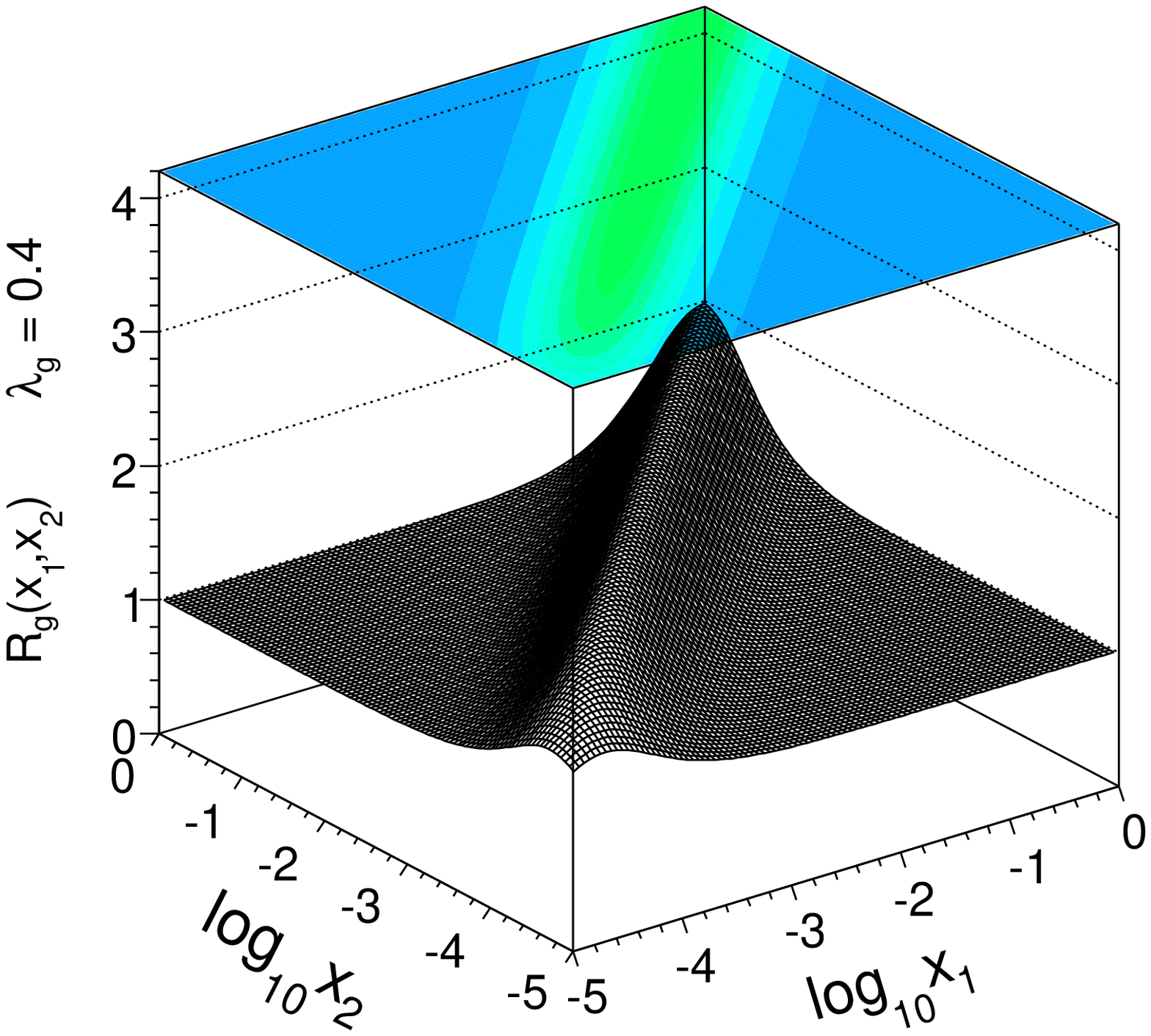} \\
 \includegraphics[width=6.0cm]{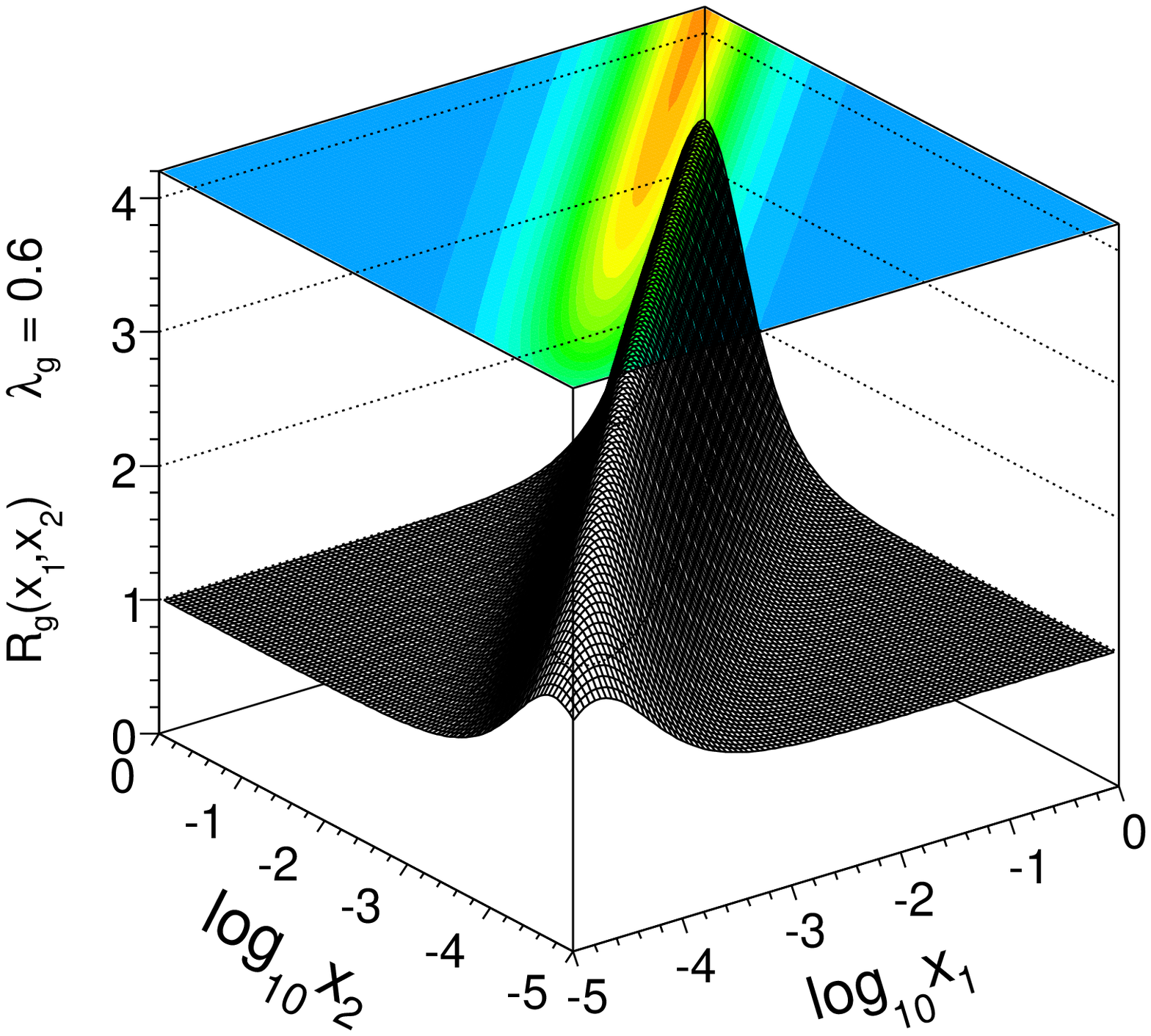}
 \includegraphics[width=6.0cm]{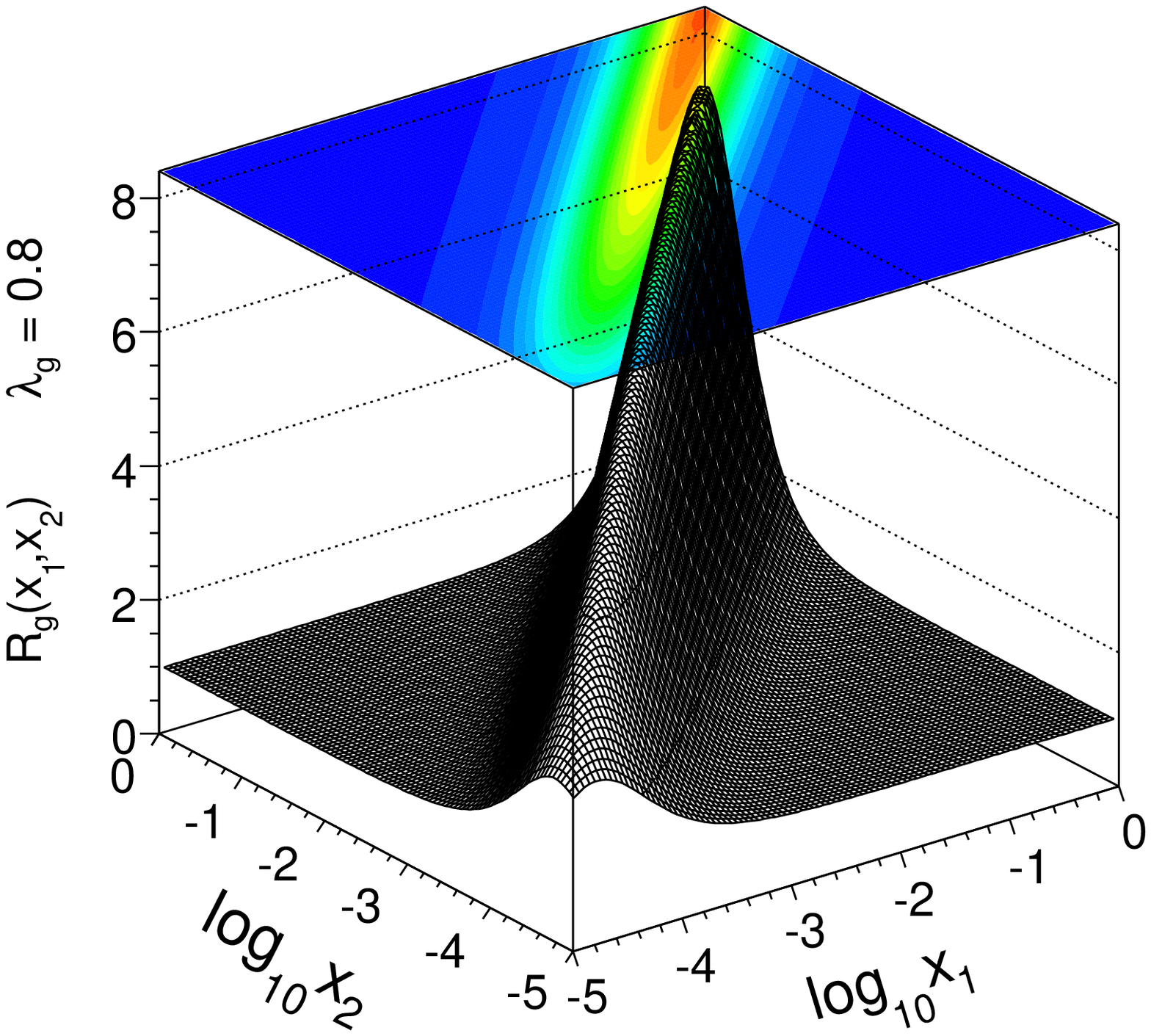}
   \caption{
\small
The ratio from Eq.(\ref{ratio}) as a function of $x_1$ and $x_2$
for different $\lambda_g$ = 0.2, 0.4, 0.6, 0.8.
}
\label{fig:ratio_lambda_g}
\end{figure}

In Fig. \ref{fig:ratio_lambda_g} we show the ratio $R_{coll}$ for different
values of $\lambda_g$. The ratio strongly depends on the value.
We observe a strong enhancement on the diagonal. The larger $\lambda_g$
the stronger the ratio. The ratio quickly drops off diagonal.
This have consequences for rapidity distributions of jets, in particular
their correlations, as is discussed in the next section.

\section{Results}

Let us start presentation of our results. Having the amplitude
${\cal M}$ for the $pp \to p(jj)p$ where $j=g,q(\bar{q})$, defined
in Eqs.~(\ref{ampl-a+b}), (\ref{ampl-c}) and (\ref{ampl}), we can
calculate the corresponding $2\to4$ cross section as
\begin{equation}
\sigma_{pp \to pjjp} = \int \frac{d^3 p_1}{(2 \pi)^3} \frac{d^3
p_2}{(2 \pi)^3}
     \frac{d^3 p_3}{(2 \pi)^3}  \frac{d^3 p_4}{(2 \pi)^3}
(2 \pi)^4 \; \delta^{(4)}(p_1 + p_2 + p_3 + p_4 - p_a - p_b) \;
\overline{ |{\cal M}_{jj}|^2 } \; . \label{cross_section}
\end{equation}
In what follows, we adopt a convenient choice of the phase space
variables of the integration relevant for exclusive diffractive
processes elaborated in Ref.~\cite{LS2010}.

\begin{figure}[!h]
\begin{minipage}{0.47\textwidth}
 \centerline{\includegraphics[width=1.0\textwidth]{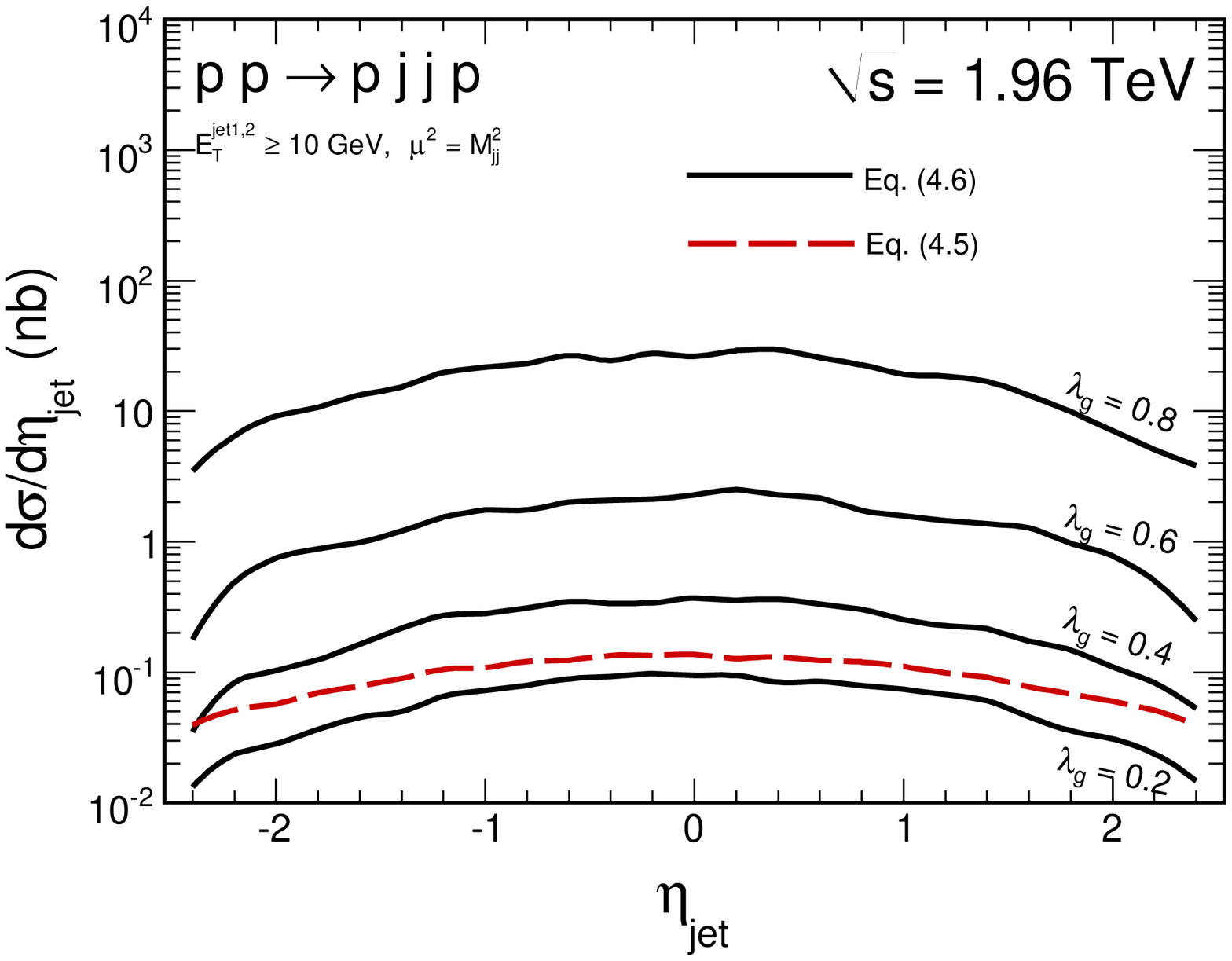}}
\end{minipage}
\hspace{0.5cm}
\begin{minipage}{0.47\textwidth}
 \centerline{\includegraphics[width=1.0\textwidth]{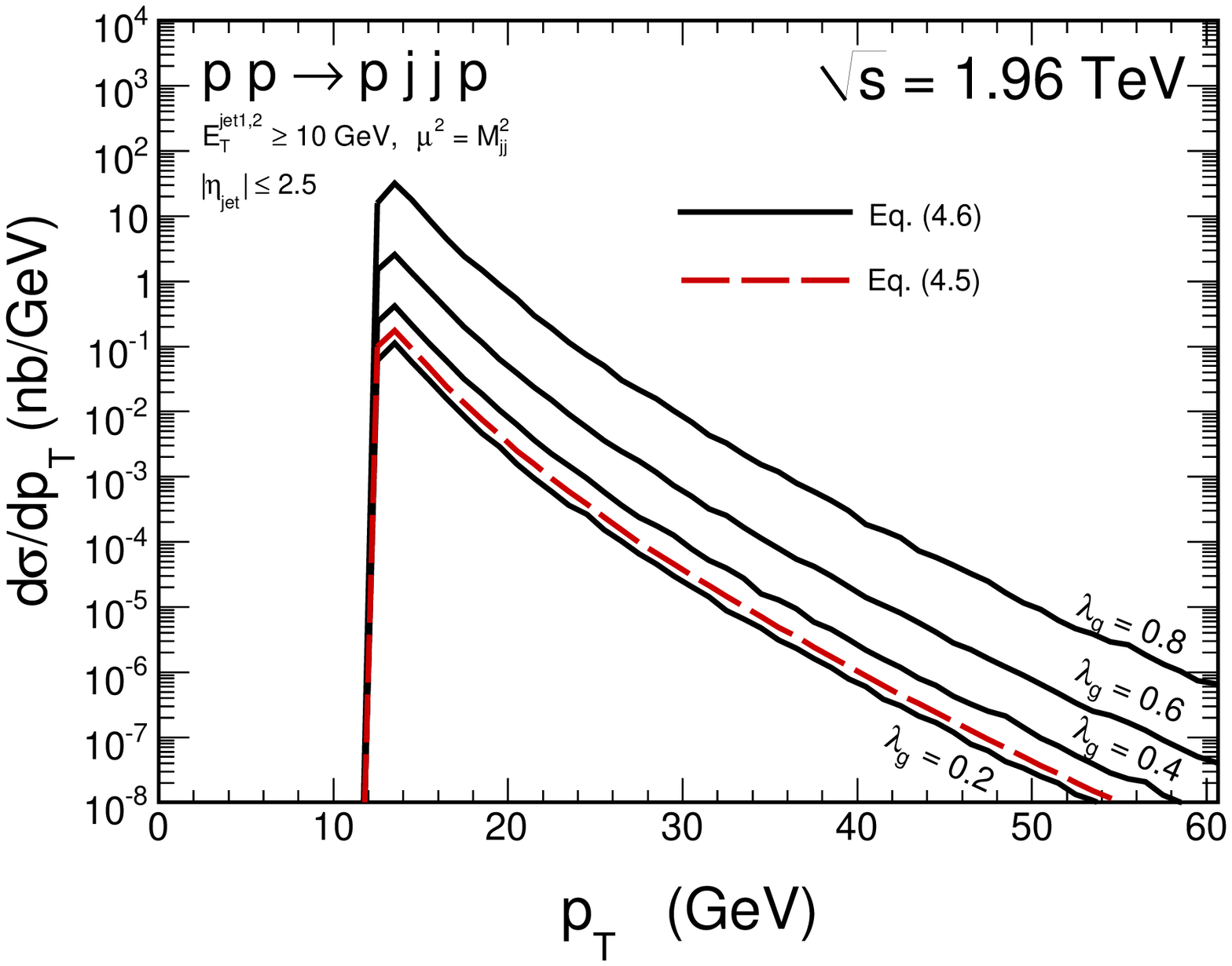}}
\end{minipage}
   \caption{
\small Distribution in pseudorapidity (left panel) and transverse momentum (right panel)
of gluonic jets for the contribution of diagram B.
We present results for different $\lambda_g$ = 0.2, 0.4, 0.6, 0.8.
The results strongly depend on the value of $\lambda_g$.
For comparison we show result (dashed line) obtained from the naive
prescription (\ref{off-diagonal-UGDFs-diagramb}).
}
 \label{fig:single_distributions_erbl}
\end{figure}

Before we go to the description of experimental data and presentation
of all contributions let us concentrate for a while on the contribution
of the diagram B mechanism of hard digluon production.
In Fig. \ref{fig:single_distributions_erbl} we show
distributions of the gluonic jets in pseudorapidity and transverse momentum
of the jet. The results have been performed for different values
of $\lambda_g$ using formula (\ref{new_prescription}) with
$R_{coll}$ based on $H_g$ from formula (\ref{small_xxi_H_g}).
The rapidity distribution strongly depends on the value of $\lambda_g$.
The dependence is stronger for smaller transverse momenta, i.e. smaller
$x$'s.

\begin{figure}[!h]
\begin{minipage}{0.47\textwidth}
 \centerline{\includegraphics[width=1.0\textwidth]{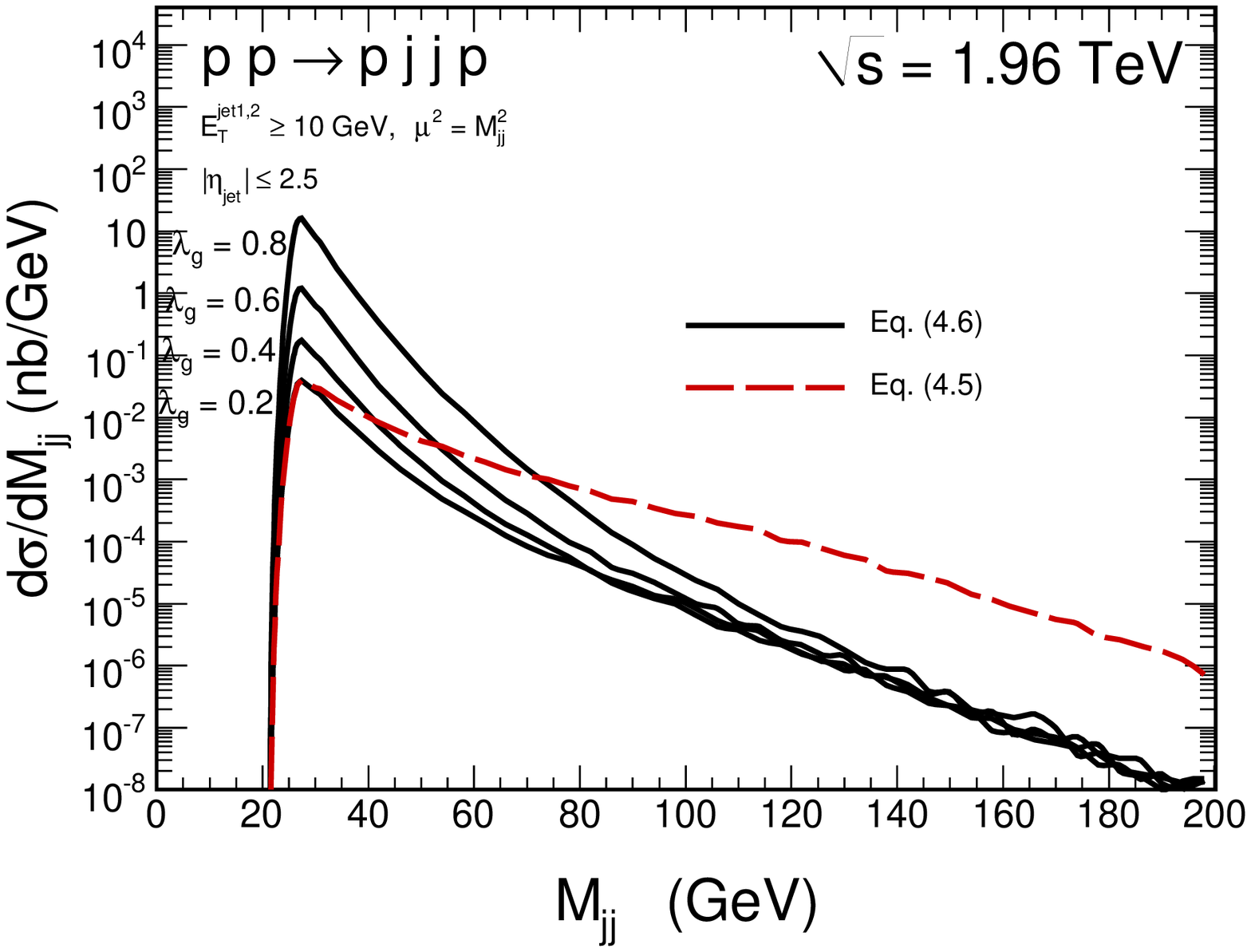}}
\end{minipage}
\hspace{0.5cm}
\begin{minipage}{0.47\textwidth}
 \centerline{\includegraphics[width=1.0\textwidth]{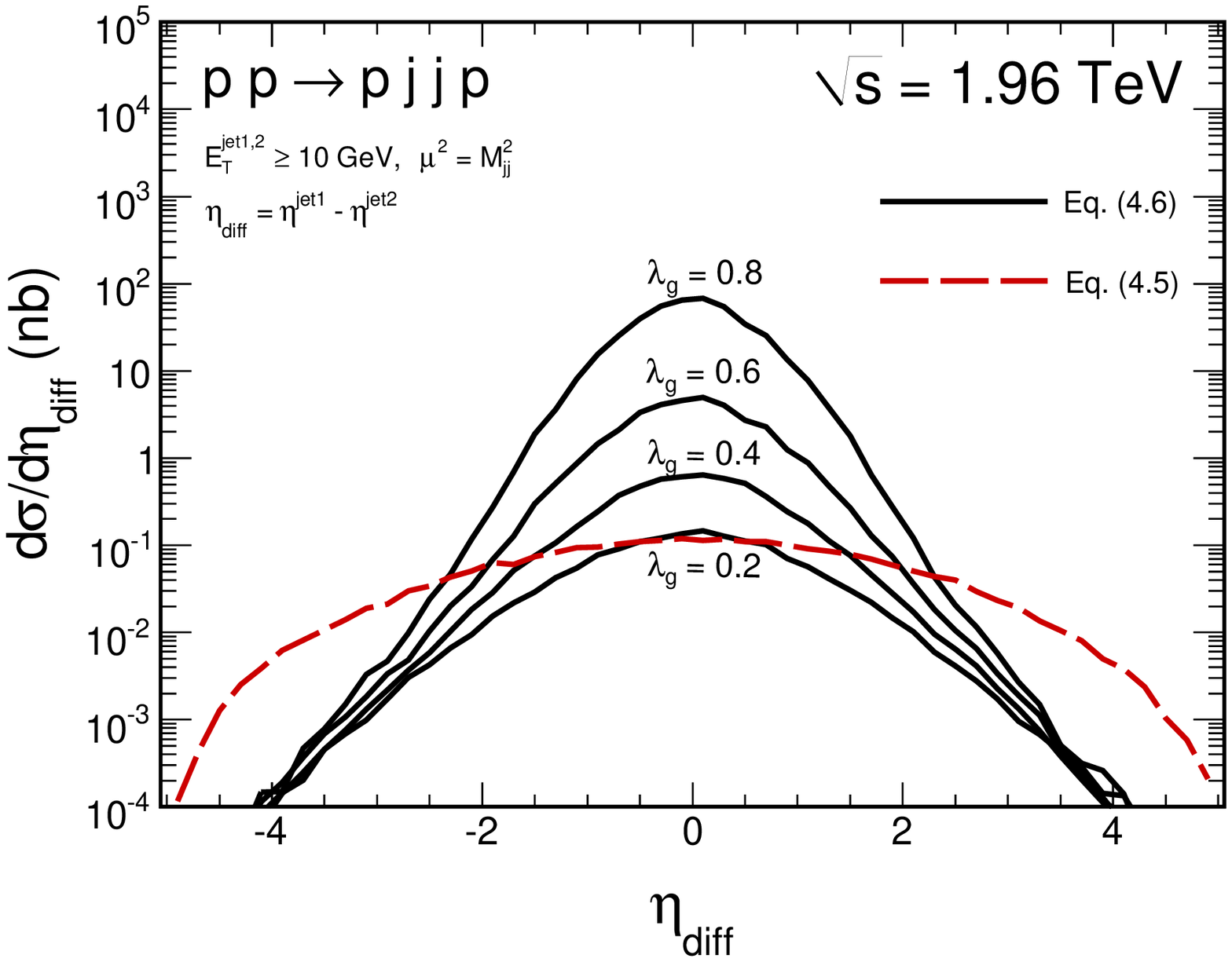}}
\end{minipage}
   \caption{
\small Distribution in dijet invariant mass (left panel) and in pseudorapidity
difference between jets (right panel) for the contribution of diagram B.
We present results for different $\lambda_g$ = 0.2, 0.4, 0.6, 0.8.
The results strongly depend on the value of $\lambda_g$.
For comparison we show result (dashed line) obtained from the naive
prescription (\ref{off-diagonal-UGDFs-diagramb}).
}
 \label{fig:dsig_dM_erbl}
\end{figure}

In Fig.\ref{fig:dsig_dM_erbl} we show similar distributions for the
dijet invariant mass (left panel). There is a stronger dependence on $\lambda_g$
at small invariant masses. In the region of the Higgs boson invariant
mass of $M_{jj}$ = 120 GeV there is a factor three uncertainties
of the cross section. For comparison we show result (dashed line)
obtained from a naive prescription (\ref{off-diagonal-UGDFs-diagramb}).
When the naive prescription gives large cross sections at large invariant masses
the improved prediction drops quickly with invariant mass. For completeness in the right panel
we show disitrbutions in rapidity difference between jets. While the distribution obtained with naive prescription
extends up to large $\eta_{diff}$, the distribution obtained with improved calculations is concentrated at small values of $\eta_{diff}$.

\begin{figure}[!h]
\begin{minipage}{0.47\textwidth}
 \centerline{\includegraphics[width=1.0\textwidth]{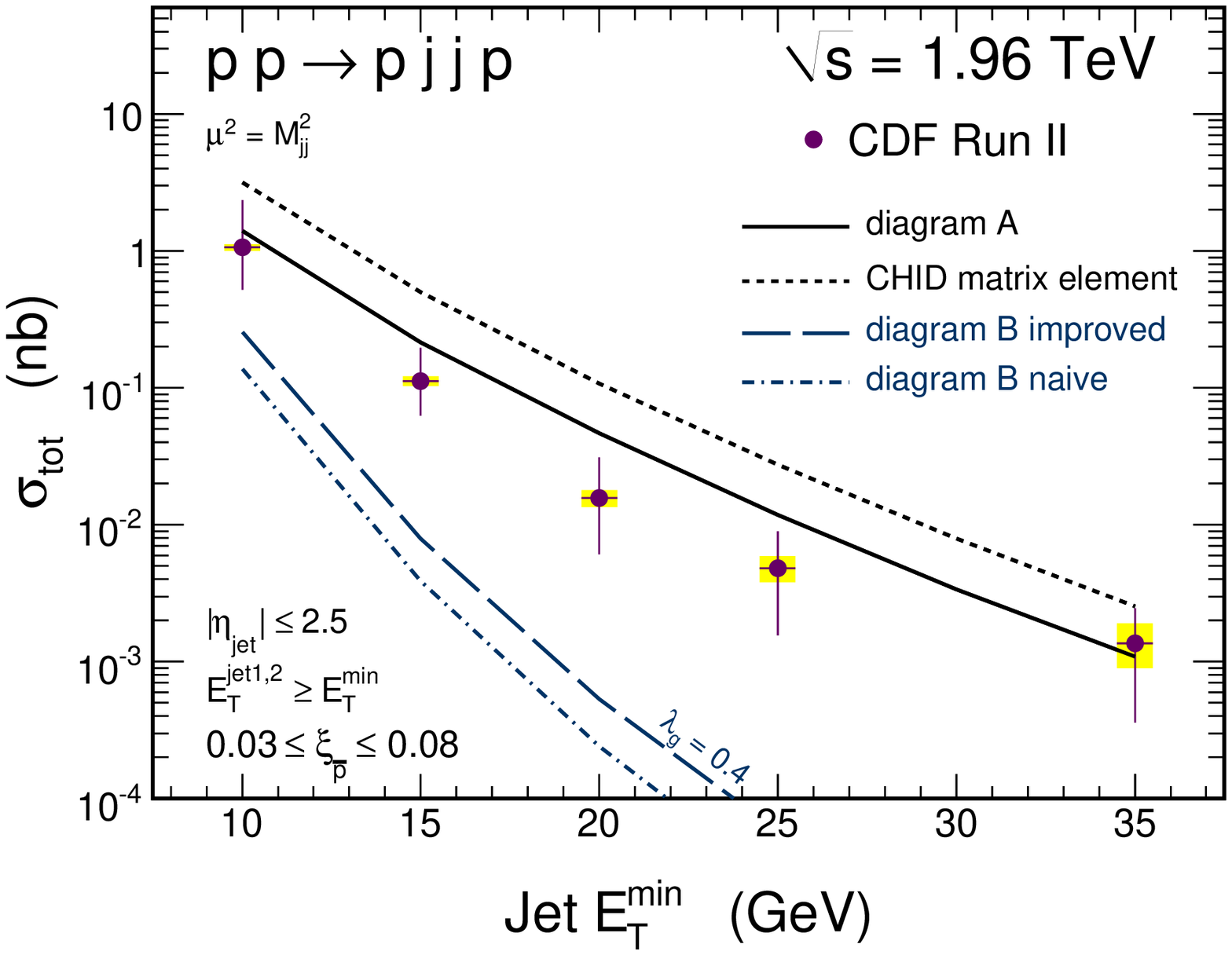}}
\end{minipage}
\hspace{0.5cm}
\begin{minipage}{0.47\textwidth}
 \centerline{\includegraphics[width=1.0\textwidth]{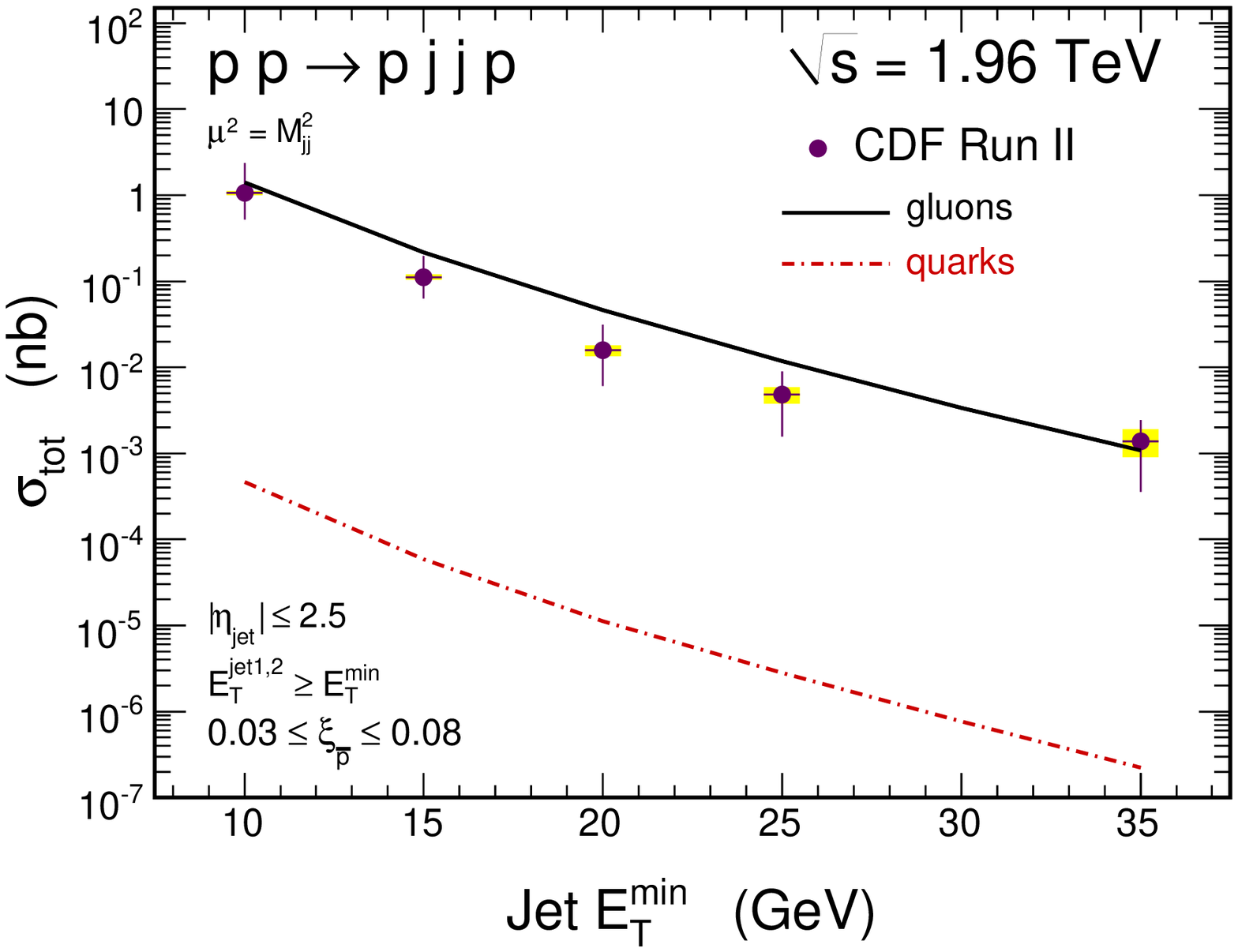}}
\end{minipage}
   \caption{
\small The total cross section as a function of $E_{t,min}$. The
experimental data points are taken from Ref.~\cite{CDF-dijets}. Left
panel: digluon contribution for diagam A with our matrex element
(solid line) and CHID matrix element (short-dashed line), for
diagram B (long-dashed line). Right panel: quark-antiquark
(dash-dotted line) contribution. }
 \label{fig:cdf_etmin}
\end{figure}

Now we shall include and discuss all contributions.
First we shall discuss numerical results obtained at the Tevatron energy
and we wish to compare our results to the existing CDF collaboration
data \cite{CDF-dijets}. This means that in the following we include
the CDF experimental cuts. In Fig.~\ref{fig:cdf_etmin} we show the
integrated cross section as a function of lower cut on $E_T$.
Following Ref.~\cite{Cudell:2008gv}, we assume the relation between
$E_T$ and the jet transverse momentum $p_{\perp}$ as $E_T$ = 0.8
$p_{\perp}$. This approximate relation could be checked in the
future by performing full simulation of jets including
hadronization. We show results for the digluon diagram A (solid
curves), as well as for the diagram B (left panel, long dashed curve)
and for the quark-antiquark jets (right panel, dash-dotted line).
The contribution of the diagram A is somewhat bigger than presented in
the literature in particular when using CHID matrix elements (left
panel, short dashed line). The main reason is that in the literature
(see e.g. Ref.~\cite{DKRS2011}) rather large lower cuts on
screening gluon transverse momentum $q_{\perp}$ are imposed (1 -- 2 GeV). Here we use gluon
distributions which allow to decrease the lower cut on the gluon
transverse momenta in the amplitude down to $q_{\perp}^{min}\simeq
0.4\,\GeV^2$ which is consistent with the Tevatron data on exclusive
production of $\chi_c$ data \cite{PST_chic}. Having such a low cut
in the case of dijets production, we rather overestimate the
experimental cross section.

The contribution corresponding to diagram B turns out to be much
smaller than that for diagram A known from the literature. In
addition, it falls much steeper with minimal $E_{T,min}$. In the
case of quark-antiquark dijets we present the contribution of $u
\bar u, d \bar d, s \bar s, c \bar c$ and $b \bar b$. In the first
three cases, we put the quark masses to zero, and in the last two
cases we take explicit masses known from the phenomenology (1.5 GeV
and 4.75 GeV, respectively). The sum of all quark-antiquark
contributions is shown in the right panel by the dash-dotted curve.
We conclude that the quark-antiquark jet contribution is smaller by
more than two orders of magnitude than the digluon one. However, as
shown in Ref.~\cite{MPS-higgs}, the $b \bar b$ contribution can be
essential e.g. as a background for Higgs searches in exclusive $pp$
scattering.

The exceptional dominance of digluon jets over quark-antiquark jets
found here offers extraordinary conditions for increased glueball
production in gluon fragmentation \cite{Ochs}. In order to
investigate it more one needs to study a contamination of central
diffractive components where the proportions of digluonic to
quark-antiquark jets are less favourable.

\begin{figure}[!h]
\begin{minipage}{0.47\textwidth}
 \centerline{\includegraphics[width=1.0\textwidth]{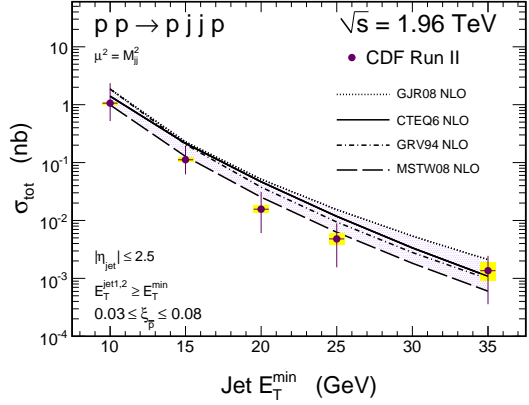}}
\end{minipage}
   \caption{
\small Uncertainties due to the choice of PDF. 
 }
 \label{fig:cdf_etmin_pdfs}
\end{figure}

\begin{figure}[!h]
\begin{minipage}{0.47\textwidth}
 \centerline{\includegraphics[width=1.0\textwidth]{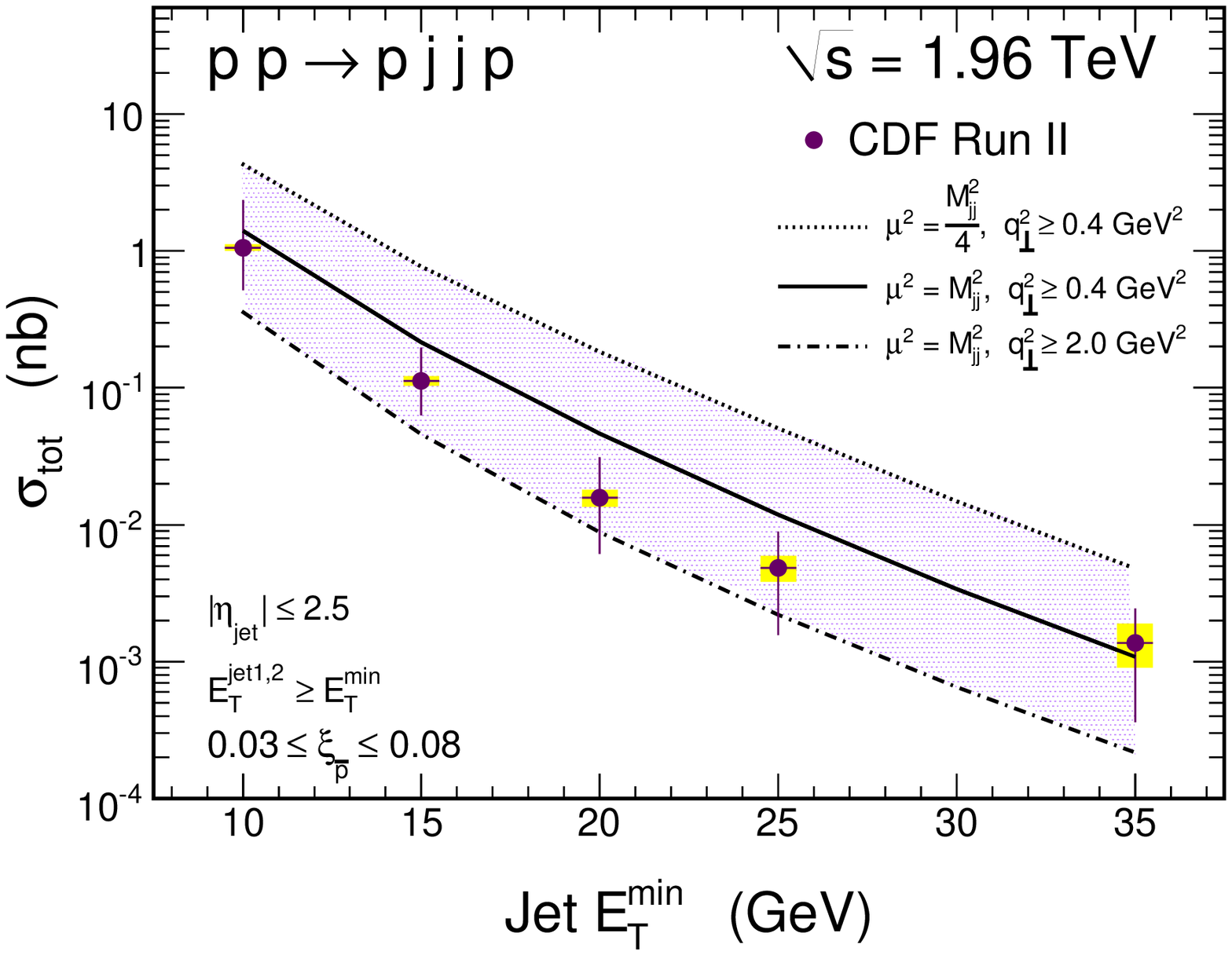}}
\end{minipage}
\hspace{0.5cm}
\begin{minipage}{0.47\textwidth}
 \centerline{\includegraphics[width=1.0\textwidth]{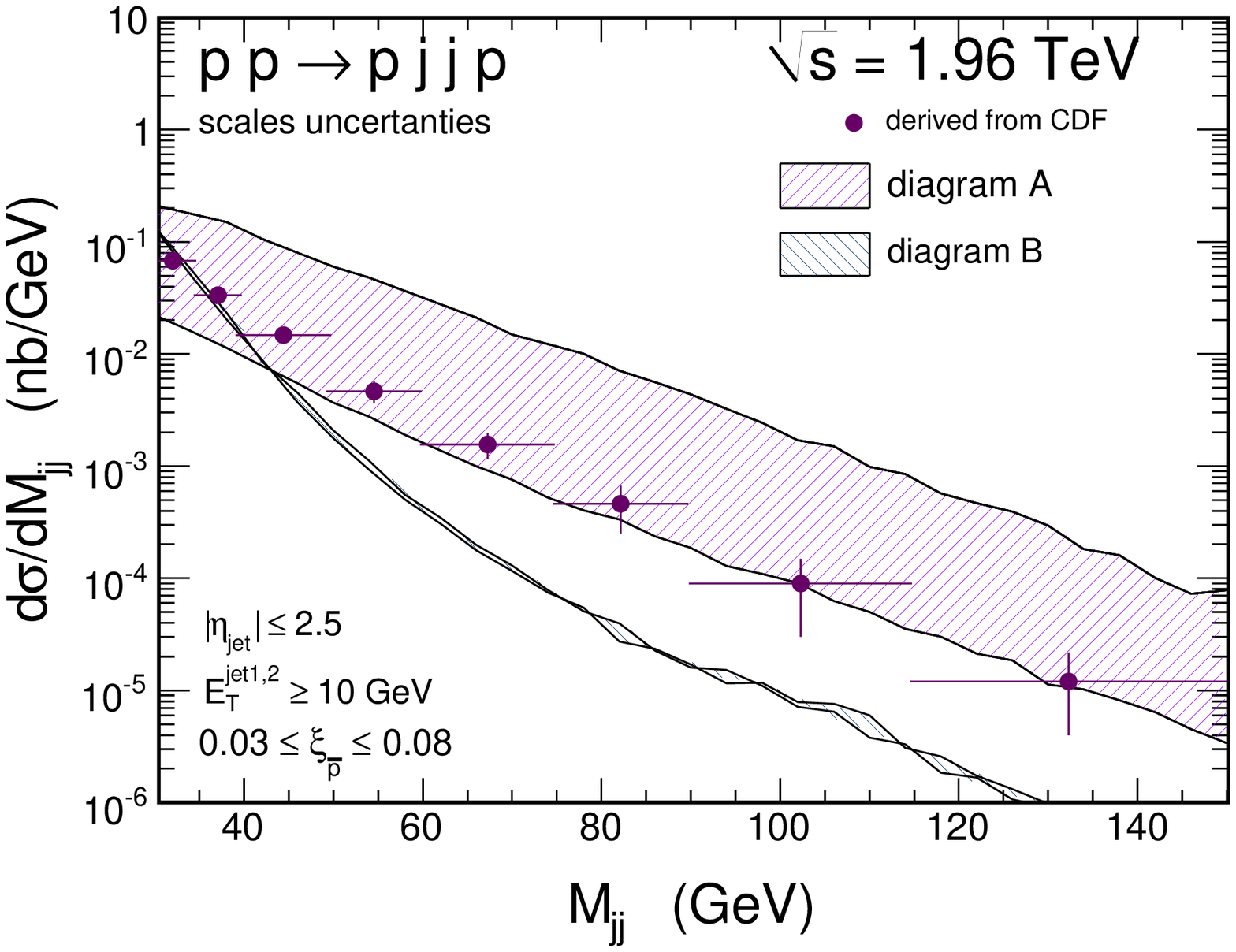}}
\end{minipage}
   \caption{
\small Uncertainties due to the choice of the scales. Left panel:
jet $E_T$ distribution for diagram A, right panel: invariant mass
distributions for both diagrams. Details are explained in the text.
 }
 \label{fig:cdf_etmin_uncert}
\end{figure}

Similarly to Ref.~\cite{DKRS2011}, we present model uncertainties
due to the choice of PDF (Fig.\ref{fig:cdf_etmin_pdfs}) and due to the
choice of the scale ($\mu_F = \mu_R$) in the left panel of
Fig.~\ref{fig:cdf_etmin_uncert} for jet $E_T^{min}$ dependence and
for dijet mass distribution in the right panel where we show
separately uncertainties for diagam A and B. We observe that they
are much smaller for diagram B. In the latter case the lower curve
corresponds to $\mu^2 = M_{jj}^2$ and the upper curve corresponds to
$\mu^2 = (p_{3\perp}^2 + p_{4\perp}^2)/2$. The smaller uncertainty
for diagram B can be explained as follows. The gluon propagators
cause that in a typical situation when two gluons coming from the
same proton line are hard and the other two are soft. This is
different compared to diagram A where typically all gluons are
rather soft (or semi-hard).

\begin{figure}[!h]
\begin{minipage}{0.47\textwidth}
 \centerline{\includegraphics[width=1.0\textwidth]{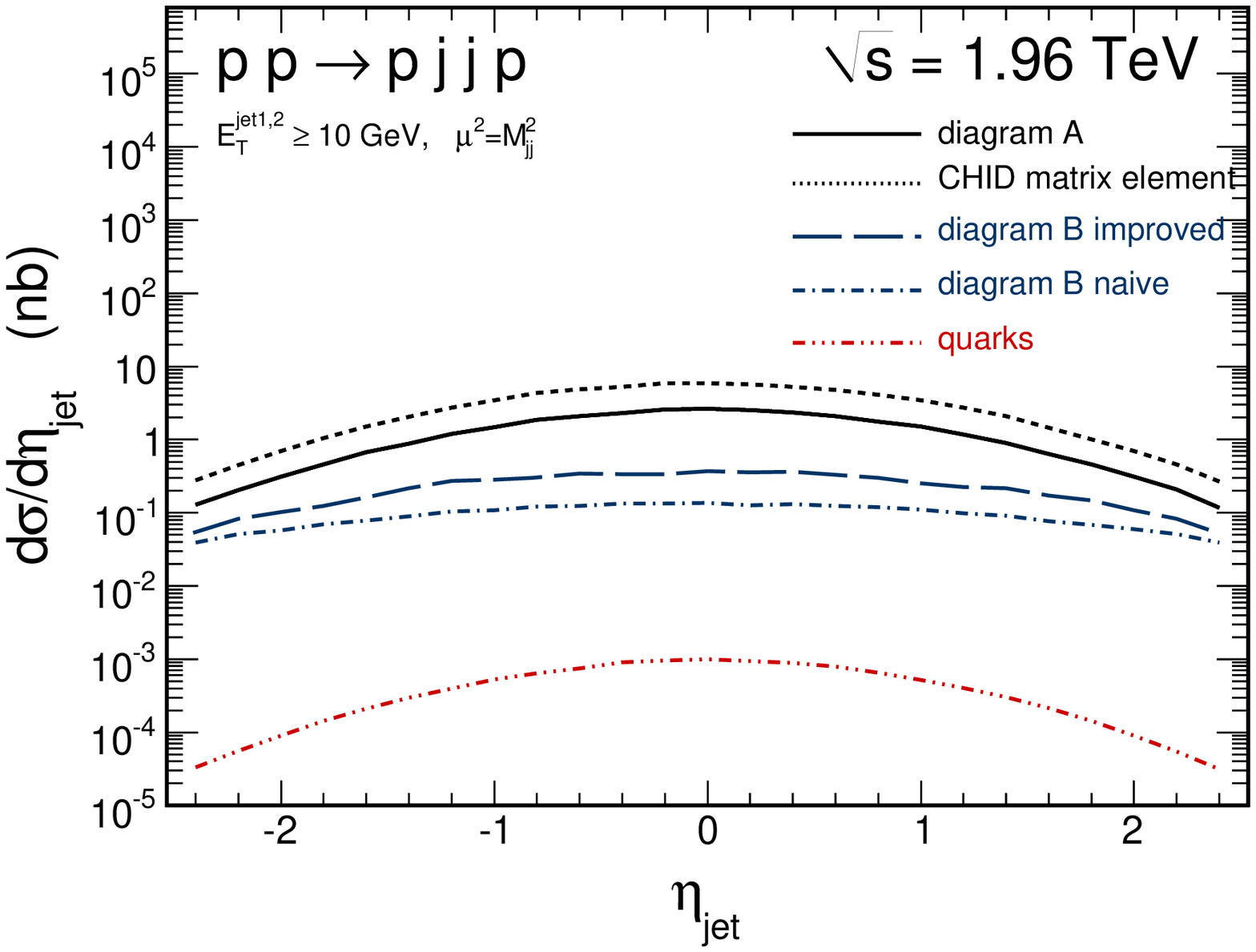}}
\end{minipage}
\hspace{0.5cm}
\begin{minipage}{0.47\textwidth}
 \centerline{\includegraphics[width=1.0\textwidth]{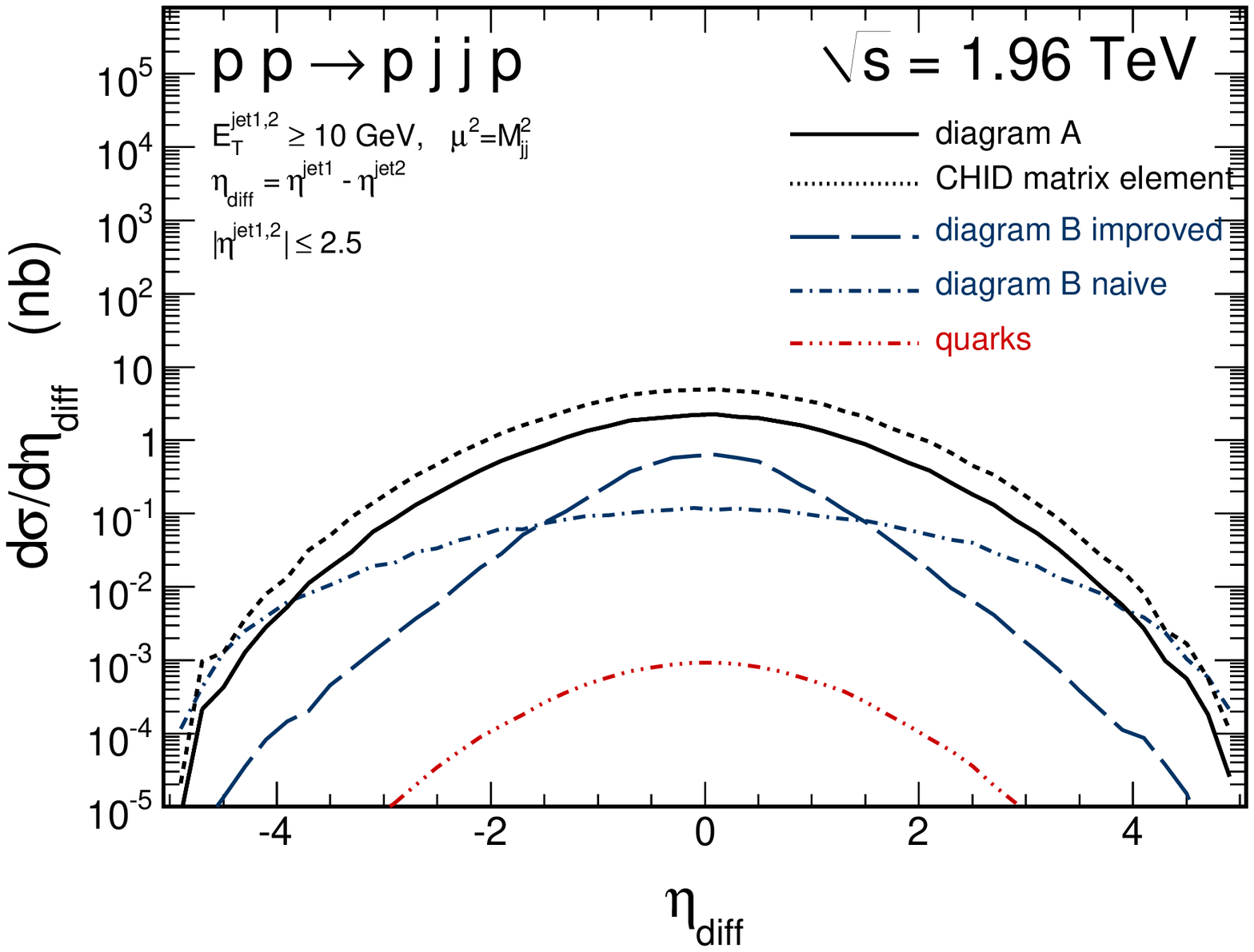}}
\end{minipage}
   \caption{
\small The distribution in jet pseudorapidity (left panel) and in
pseudorapidity difference (right panel).}
 \label{fig:rapidity-tevatron}
\end{figure}

In Fig.~\ref{fig:rapidity-tevatron} we show pseudorapidity
distributions of one of the gluonic jets (left panel) and
distribution in jet pseudorapidity difference (right panel) for the
fixed lower cut on $E_T$. The distribution for naively calculated
diagram B is flatter than that for diagram A. The same is true in
rapidity difference where the two contributions are almost identical
for large rapidity differences where, however, the cross section is
rather small. The corresponding distributions for improved method
for calculation of diagram B are quite different than those for naive
calculation. The contribution of quark-antiquark jets (dash-dotted
curve) is negligible. The rapidity distributions were not presented
by the CDF collaboration in Ref.~\cite{CDF-dijets}.

\begin{figure}[!h]
\begin{minipage}{0.47\textwidth}
 \centerline{\includegraphics[width=1.0\textwidth]{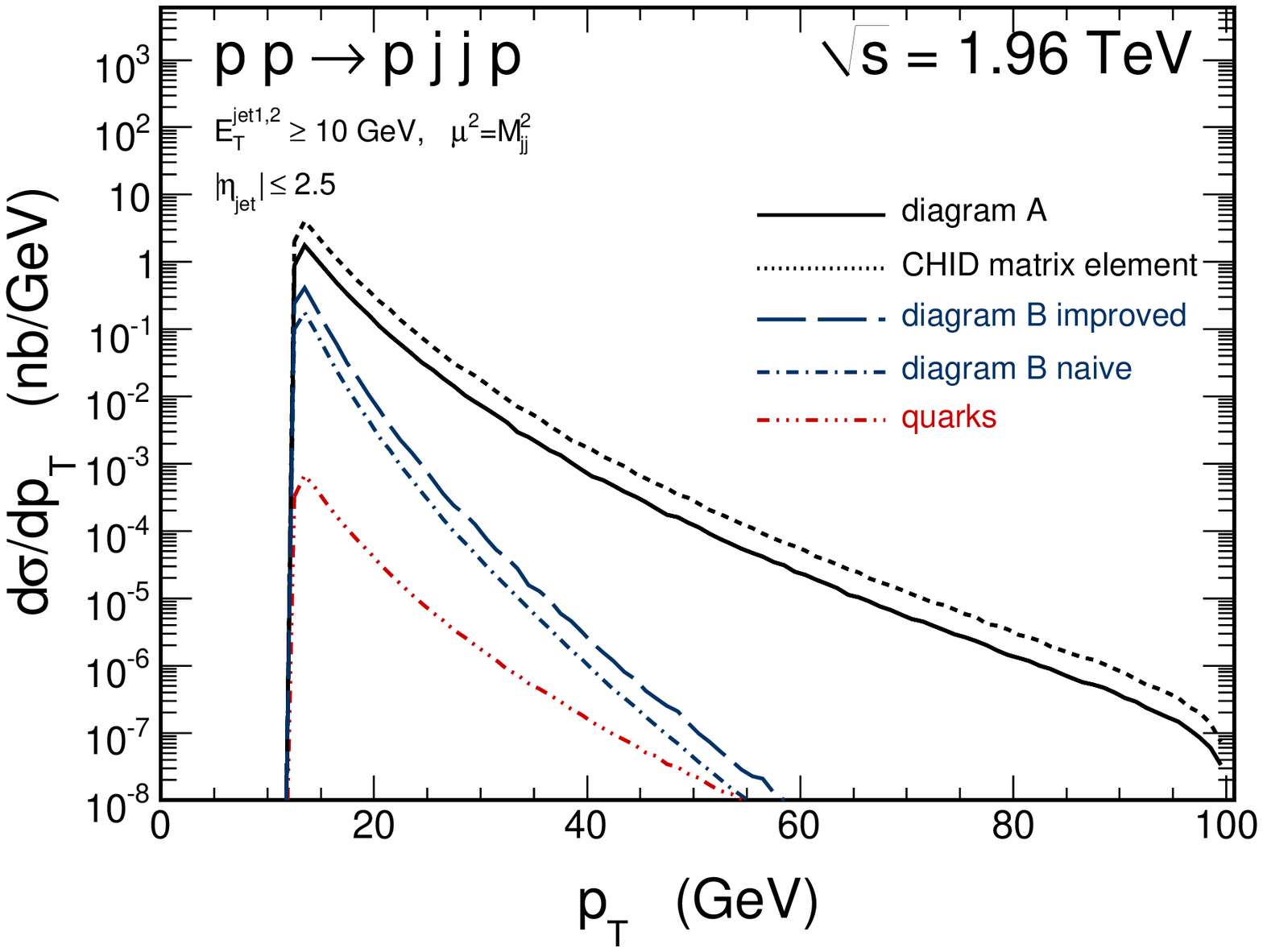}}
\end{minipage}
\hspace{0.5cm}
\begin{minipage}{0.47\textwidth}
 \centerline{\includegraphics[width=1.0\textwidth]{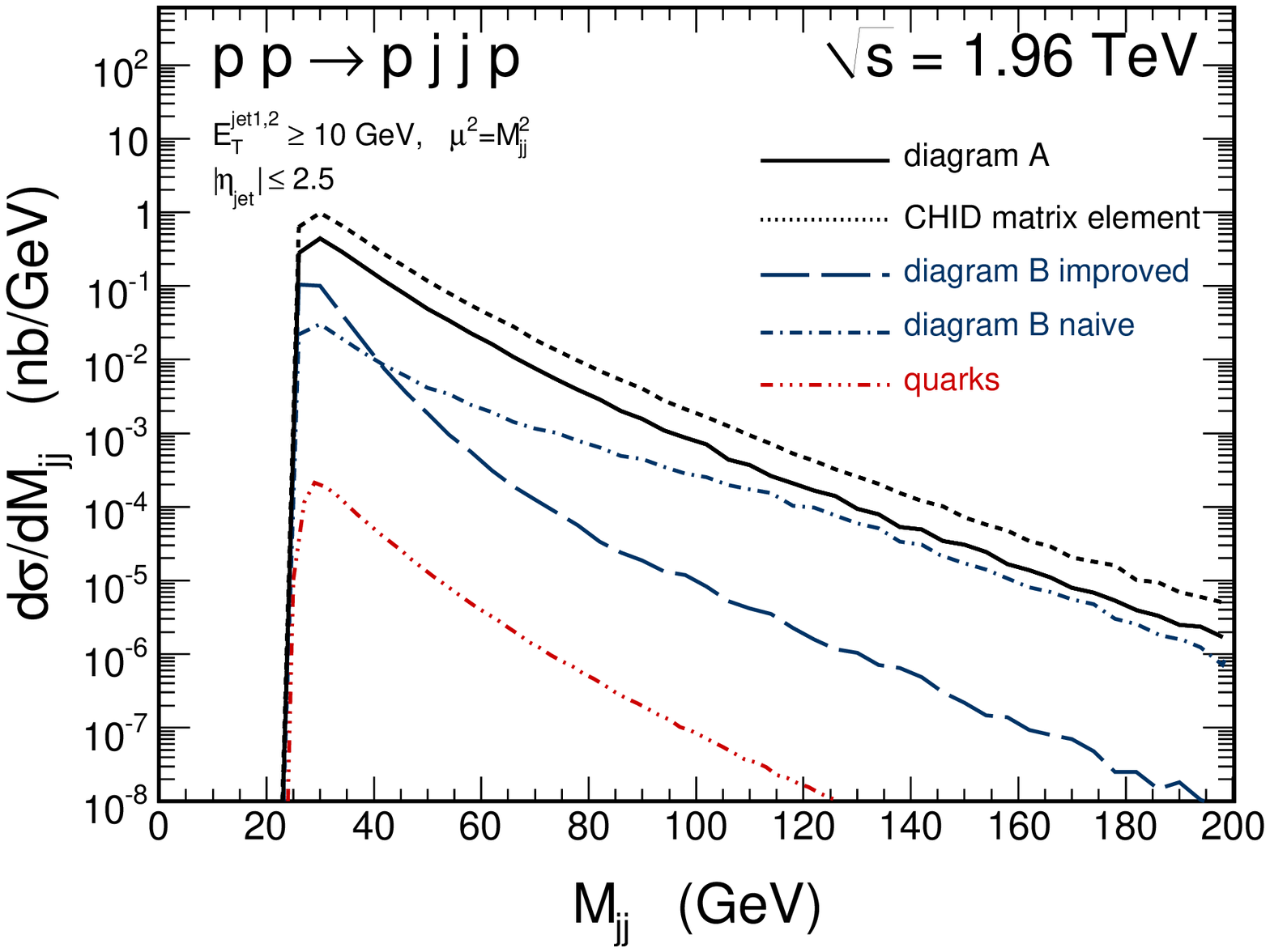}}
\end{minipage}
   \caption{
\small The distribution in jet transverse momentum (left panel) and
in dijet invariant mass (right panel). }
 \label{fig:pt-mass-tevatron}
\end{figure}

In the left panel of Fig.~\ref{fig:pt-mass-tevatron} we show in
addition the corresponding distributions in jet transverse momentum.
The contribution of diagram B becomes negligible at large jet
transverse momenta (or transverse energy). In the right panel we
show the distribution in dijet invariant mass. At large invariant
masses, naive calculations of diagram B give cross sections which
are similar to the leading contribution from diagram A. However, the
contribution of diagram B from improved prescription is sizeable
only at small invariant masses and does not have any meaning in the
important for Higgs searches large $M_{jj}$ region.

\begin{figure}[!h]
\begin{minipage}{0.47\textwidth}
 \centerline{\includegraphics[width=1.0\textwidth]{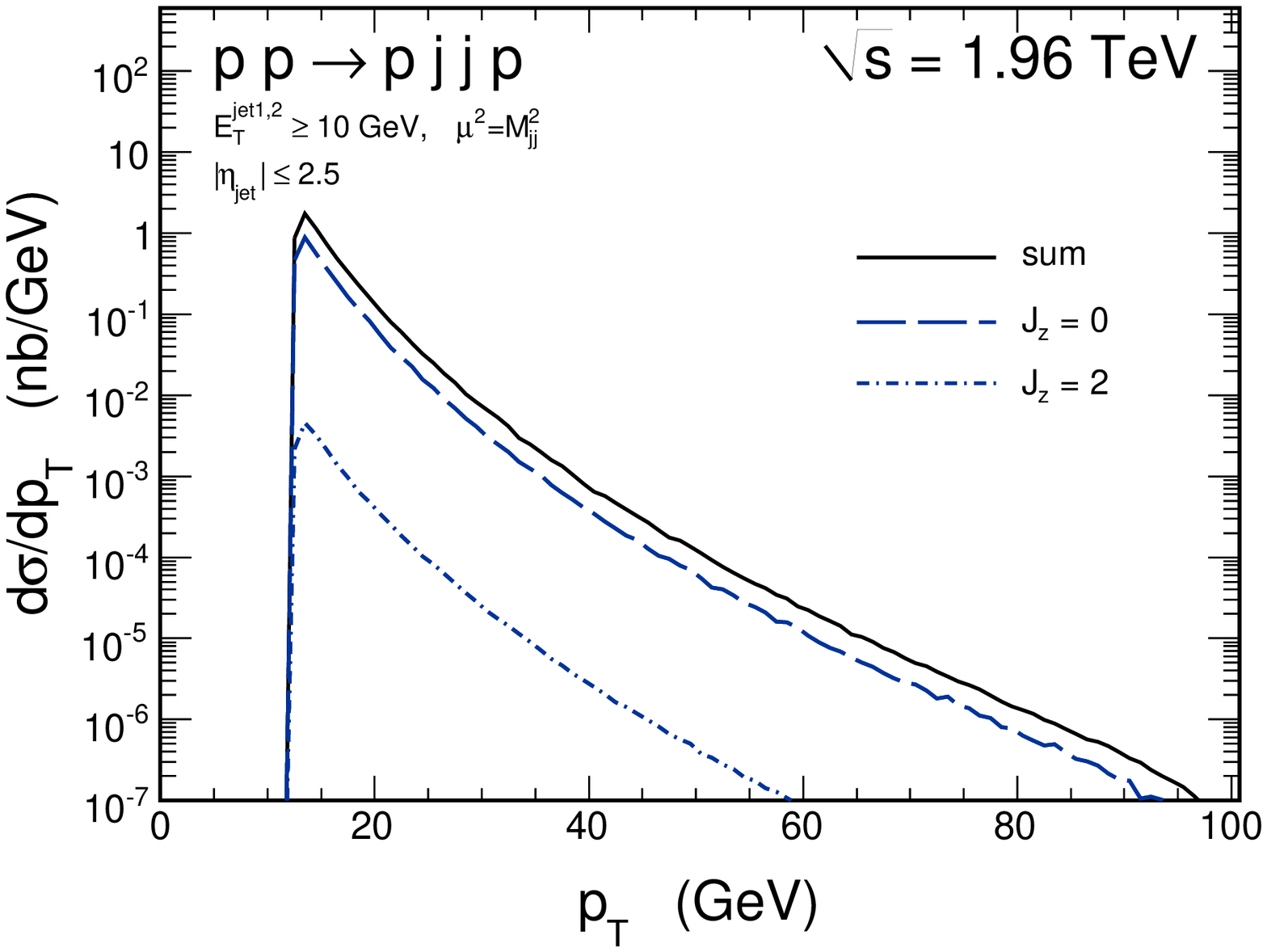}}
\end{minipage}
\hspace{0.5cm}
\begin{minipage}{0.47\textwidth}
 \centerline{\includegraphics[width=1.0\textwidth]{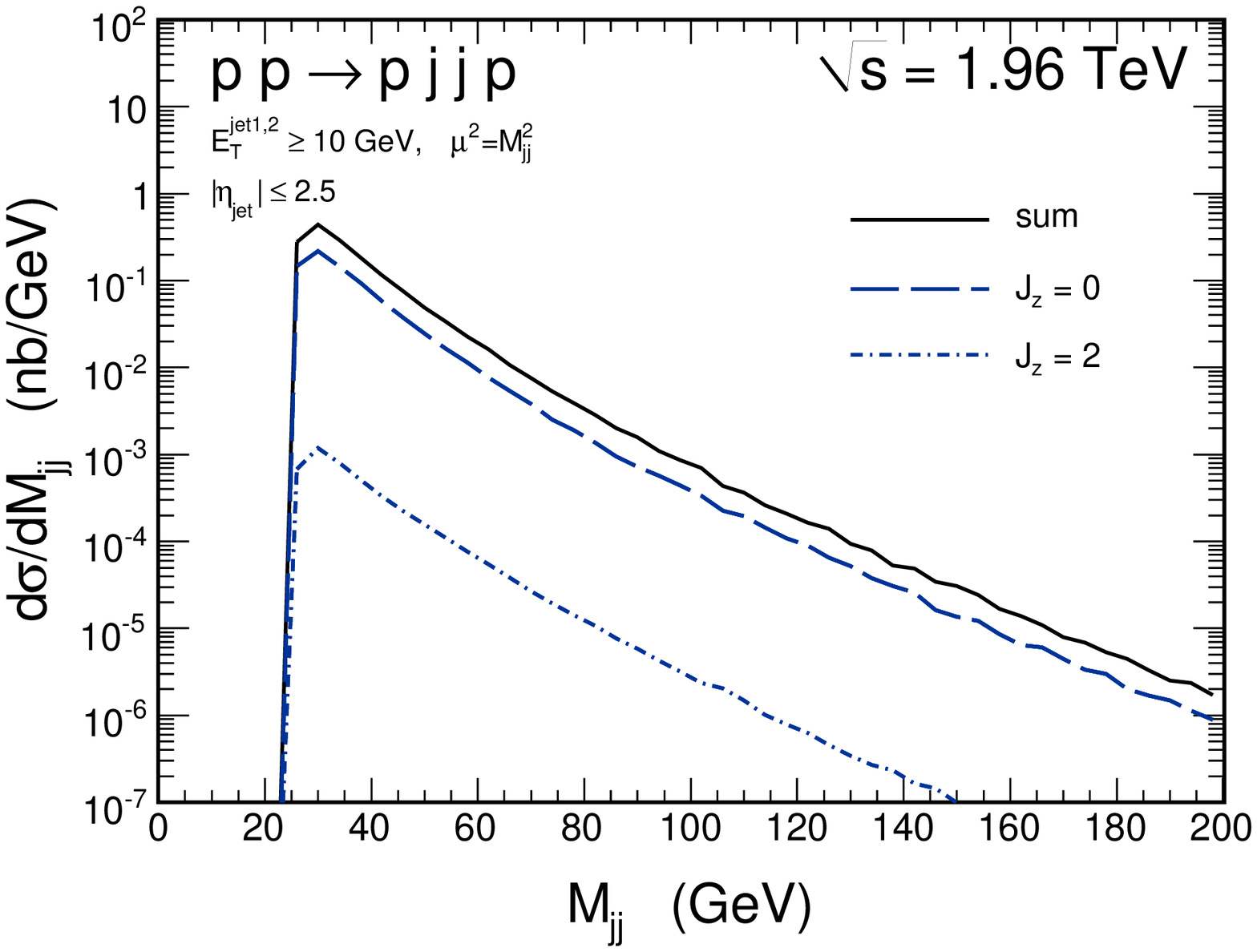}}
\end{minipage}
   \caption{
\small The distribution in jet transverse momentum (left panel) and
in dijet invariant mass (right panel) for different dijet helicity
states. Only diagram A is included here.}
 \label{fig:pt-mass-helicity-tevatron}
\end{figure}

In Fig.~\ref{fig:pt-mass-helicity-tevatron} we show separately
contributions for different helicity combinations. Being fully
consistent with the $J_z$ = 0, we see the dominance of the $++$ =
$--$ contributions over $+-$ = $-+$ ones. However, our helicities
are in the proton-proton center-of-mass system so the relation to
the $J_z$ = 0 rule is only approximate and strictly valid in the
high-$p_\perp$ jets limit.

\begin{figure}[!h]
\begin{minipage}{0.47\textwidth}
 \centerline{\includegraphics[width=1.0\textwidth]{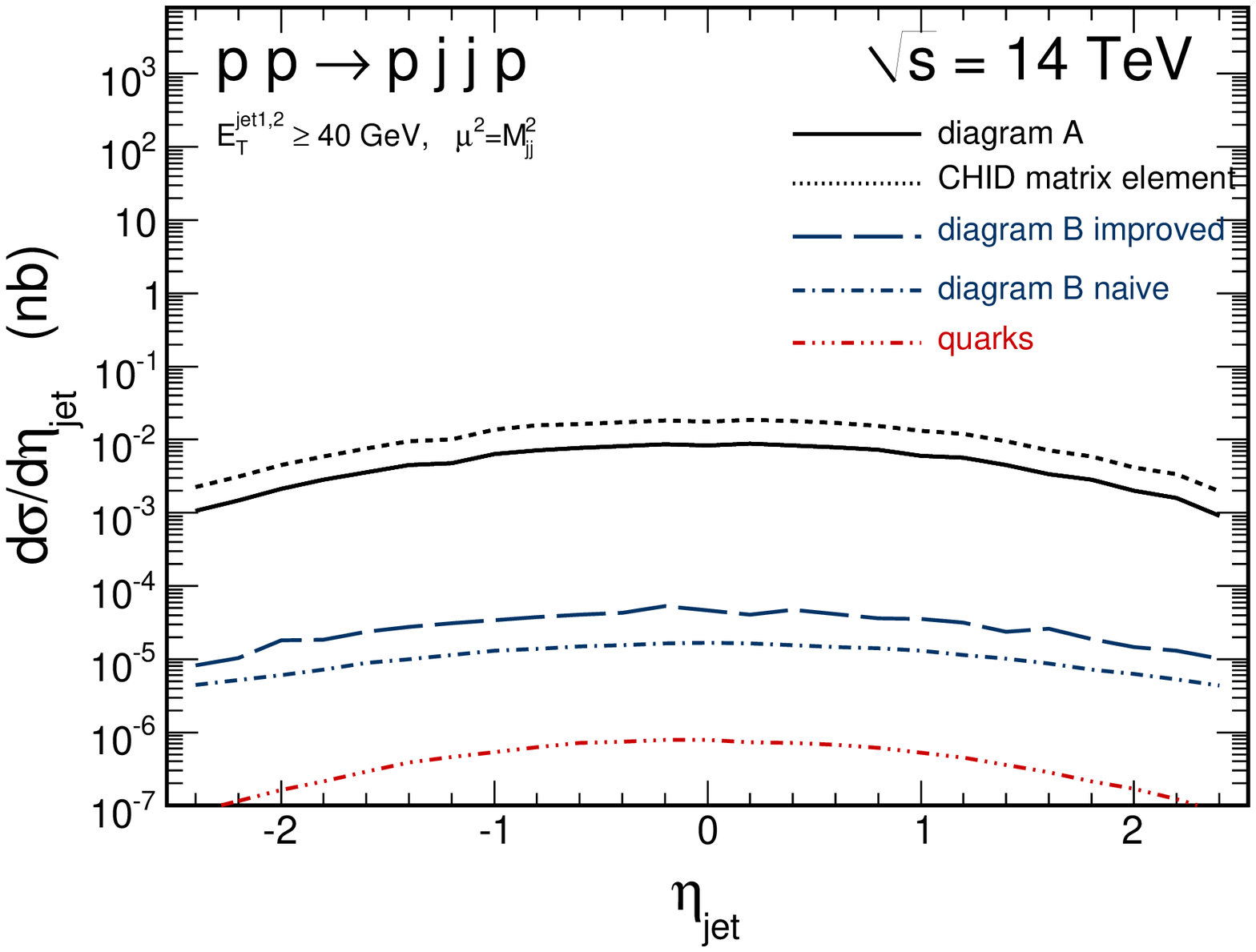}}
\end{minipage}
\hspace{0.5cm}
\begin{minipage}{0.47\textwidth}
 \centerline{\includegraphics[width=1.0\textwidth]{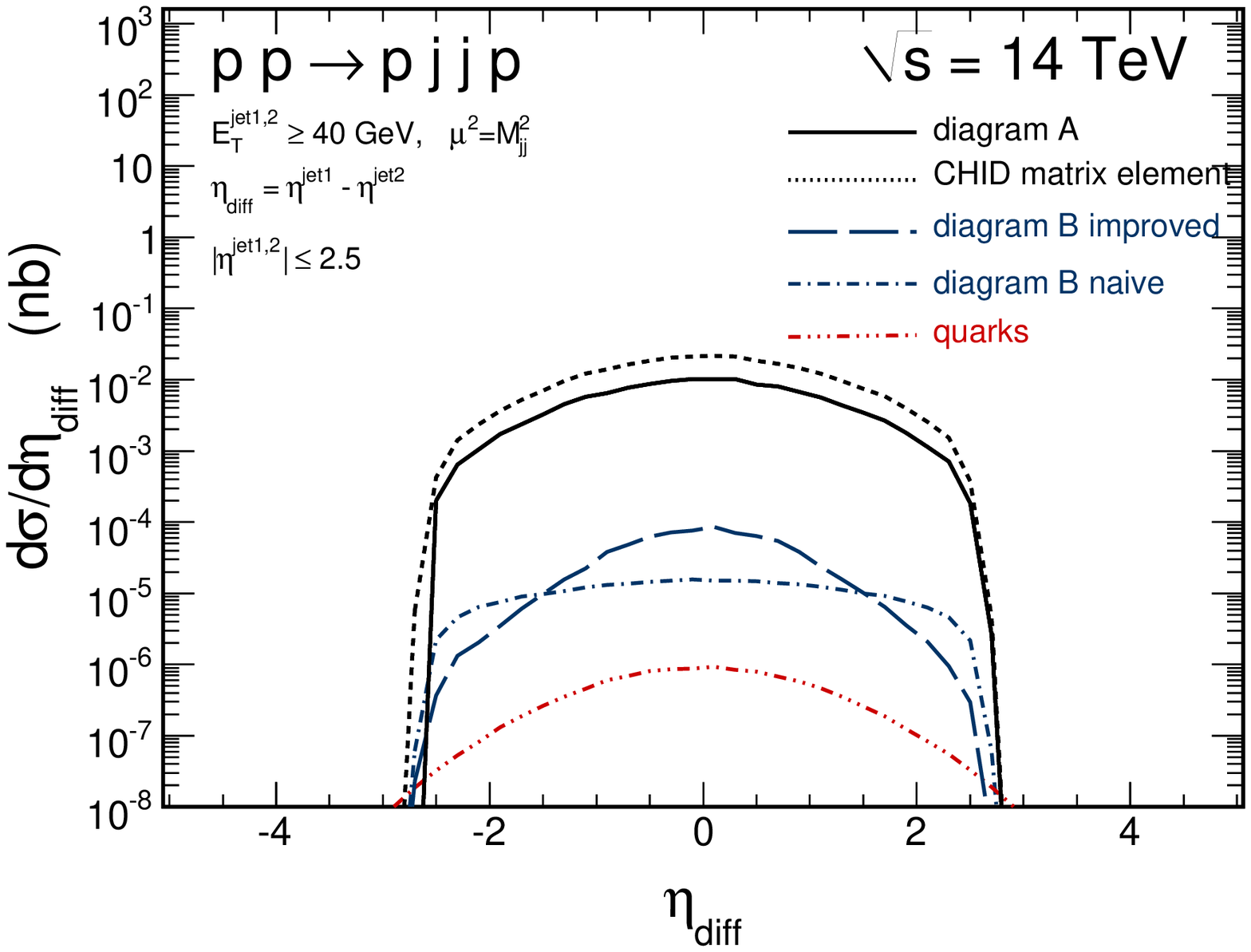}}
\end{minipage}
   \caption{
\small The distribution in jet pseudorapidity (left panel) and in
pseudorapidity difference (right panel). }
 \label{fig:rapidity-lhc}
\end{figure}

\begin{figure}[!h]
\begin{minipage}{0.47\textwidth}
 \centerline{\includegraphics[width=1.0\textwidth]{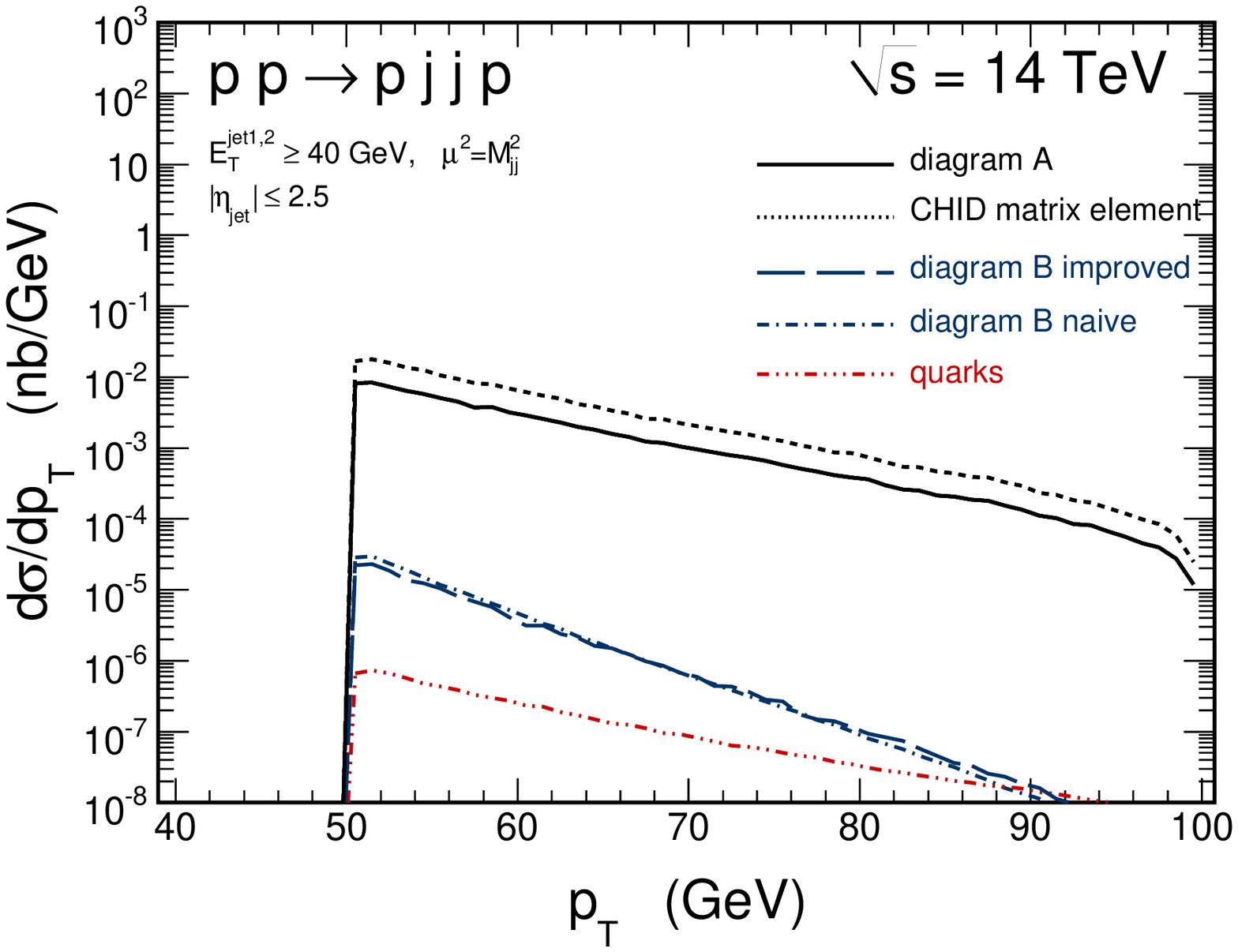}}
\end{minipage}
\hspace{0.5cm}
\begin{minipage}{0.47\textwidth}
 \centerline{\includegraphics[width=1.0\textwidth]{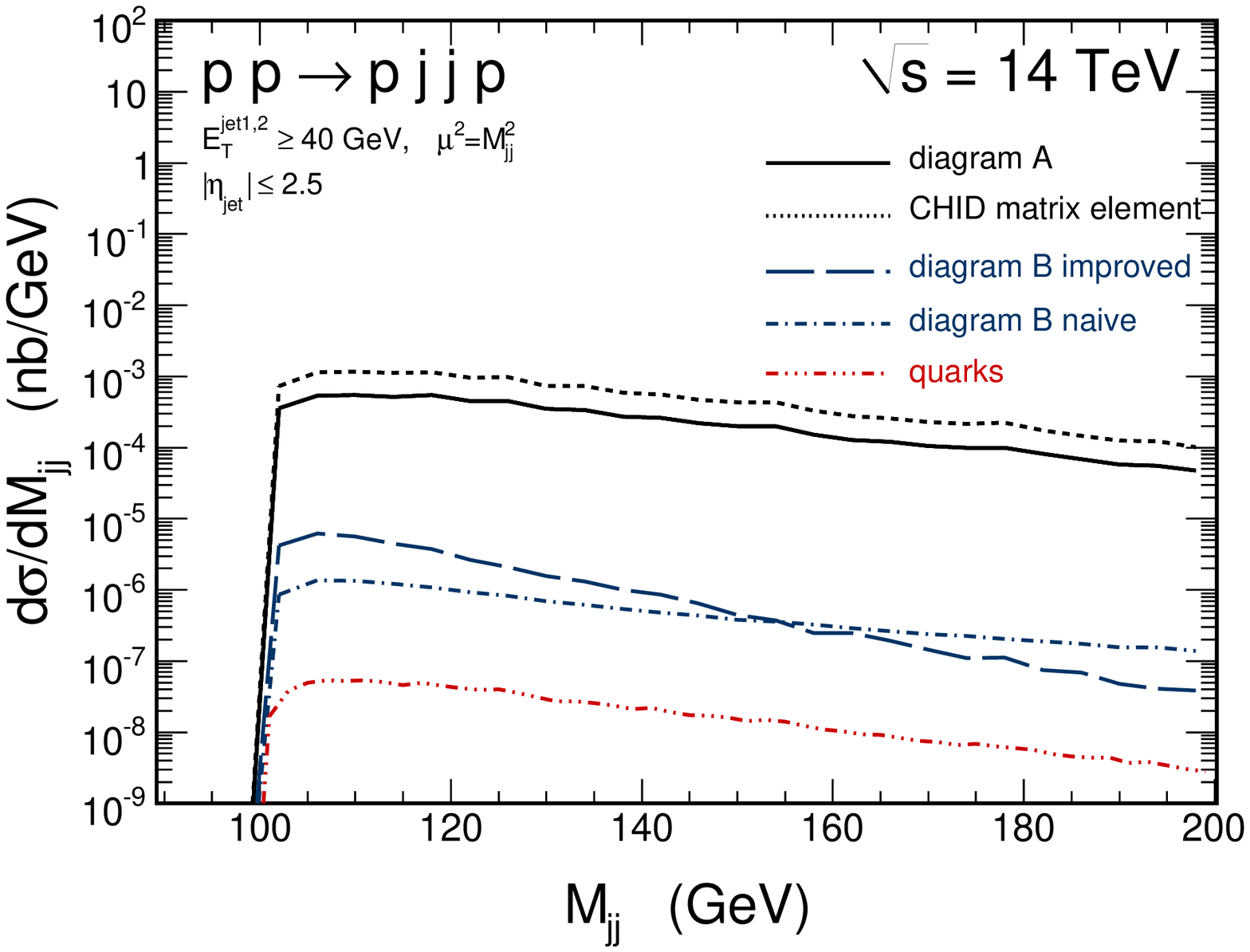}}
\end{minipage}
   \caption{
\small The distribution in jet transverse momentum (left panel) and
in dijet invariant mass (right panel). }
 \label{fig:pt-mass-lhc}
\end{figure}

We show similar distributions at nominal LHC energy $\sqrt{s}=14$
TeV in Fig.~\ref{fig:rapidity-lhc} and \ref{fig:pt-mass-lhc}. The
situation is qualitatively similar as for the Tevatron case. Here
the distributions in jet transverse momentum and dijet invariant
mass are somewhat flatter. At the LHC energy, the contribution of
diagram B is much smaller than for diagram A in the whole range of
kinematical variables considered here.

\begin{figure}[!h]
\includegraphics[width=5.0cm]{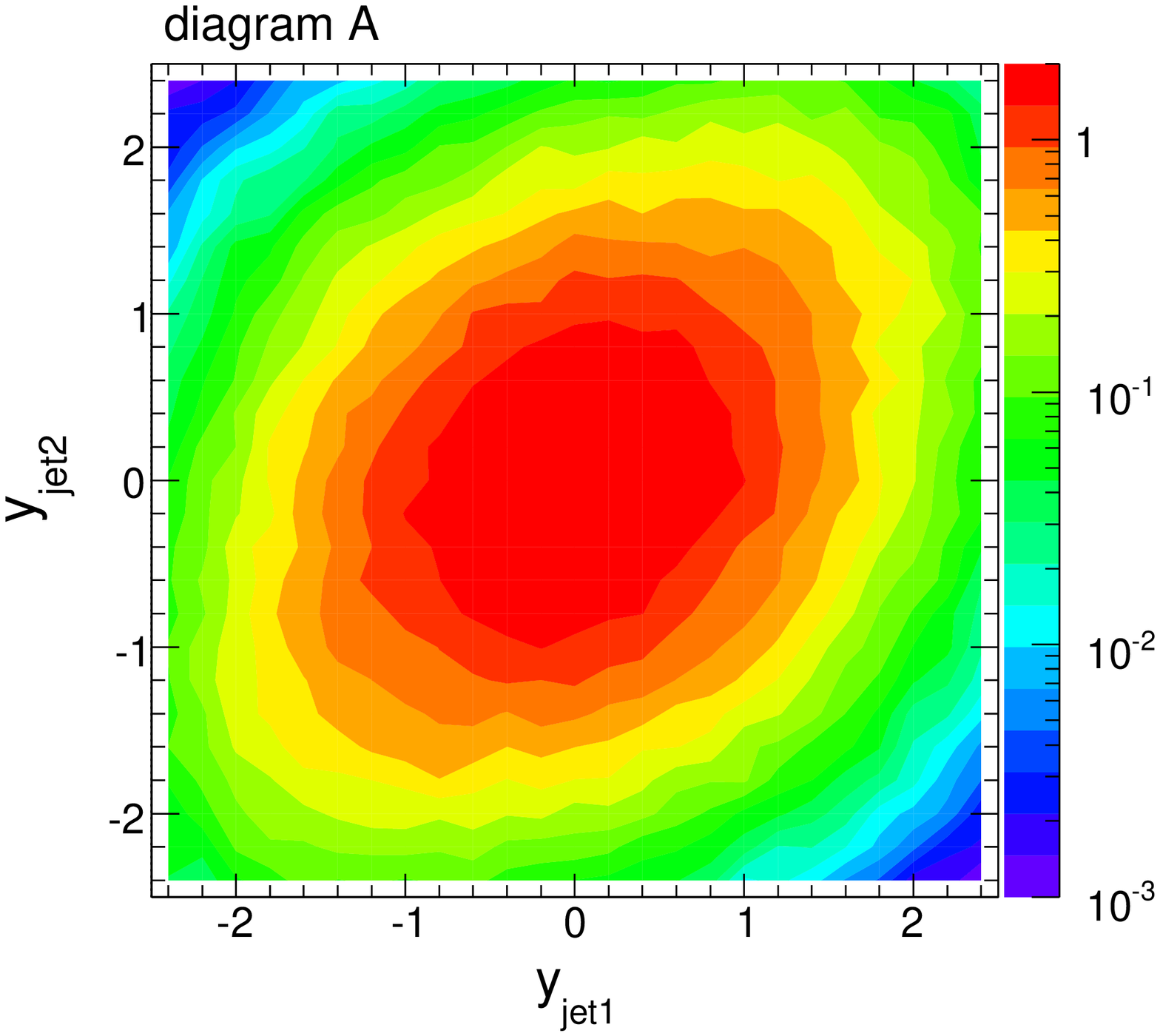}
\includegraphics[width=5.0cm]{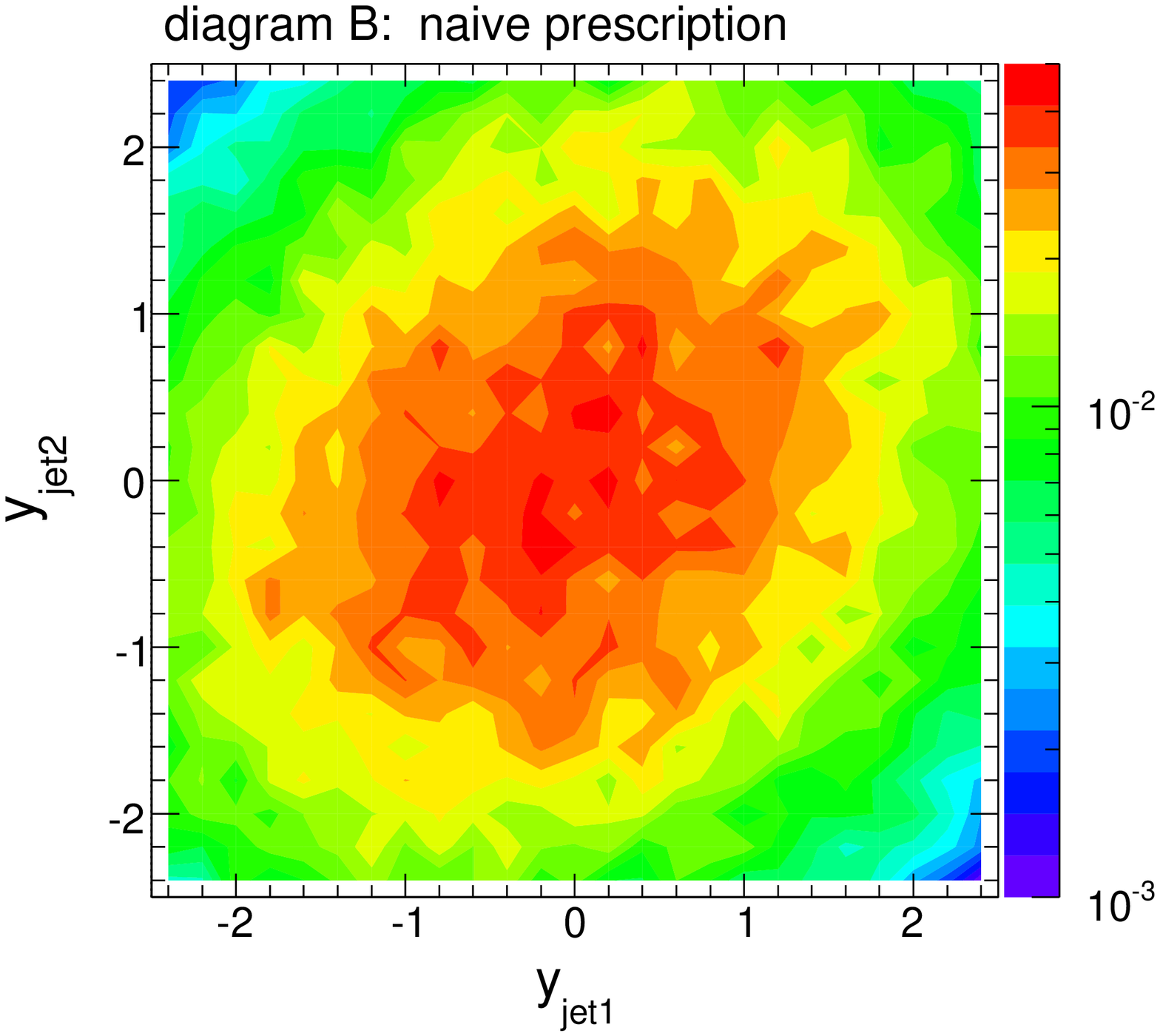}
\includegraphics[width=5.0cm]{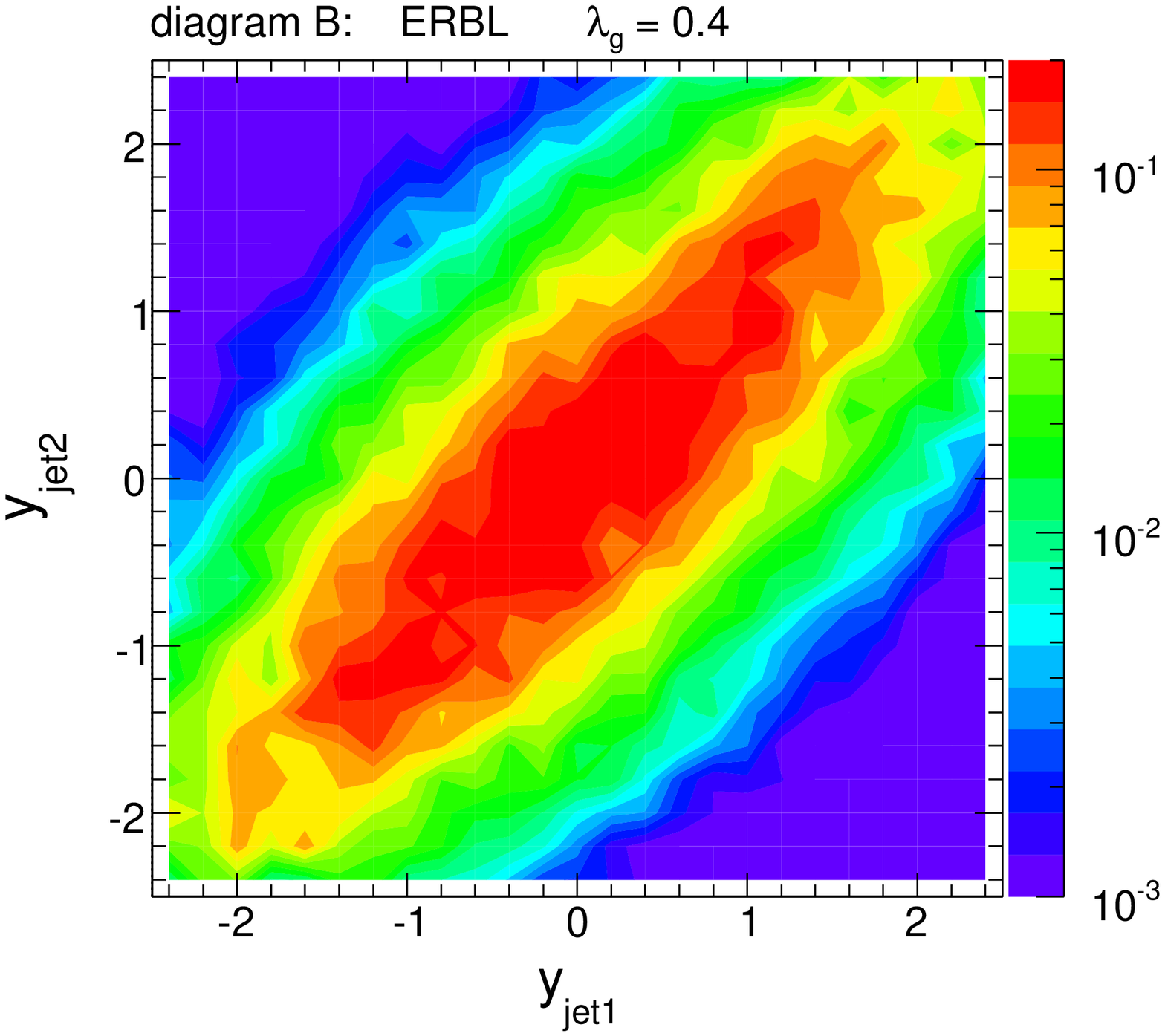}
   \caption{
\small $d\sigma / dy_3 dy_4$ for diagram A (left panel), for diagram
B, naive prescription (middle panel) and for diagram B calculated in
the way proposed in this paper (right panel).}
 \label{fig:y3y4}
\end{figure}

Finally, we wish to discuss correlations in rapidity between gluonic
jets. In Fig.\ref{fig:y3y4} we show distribution for diagram A
(left panel), naively calculated contribution of diagram B (middle panel) and
contributions of diagram B calculated as proposed above (right panel).

The correlation for the first and second case is similar. The two
gluonic jets are only weakly correlated in these variables.
It is completely different for the contribution of diagram B calculated
using the ratio of collinear off-diagonal gluon distributions
(see Eq.(\ref{ratio})). We observe a strong ridge along the diagonal
$y_3 = y_4$. This ridge is a consequence of the ratio defined in
Eq.(\ref{ratio}).

\begin{figure}[h!]
\includegraphics[width=8.0cm]{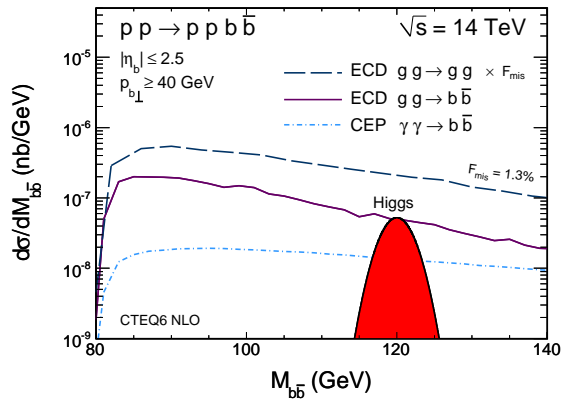}
 \caption{
Invariant mass distribution of the $b \bar b$ system. Shown are
contributions from diffractive Higgs boson (shaded area), $b \bar b$
continuum (solid line), $\gamma \gamma$ continuum (dash-dotted line) and
diffractive digluon contribution (dashed line) multiplied by an ATLAS
misidentification factor squared.
}
\label{fig:dsig_dMjj_Higgs}
\end{figure}

In Ref.\cite{MPS-bbbar} we have studied in detail irreducible exclusive
$b \bar b$ background to exclusive Higgs boson production.
The gluonic jets can be misidentified as $b$-quark jets \cite{Cox:2007sw}.
If both gluonic jets are misidentified then such a misidentified event
can contribute to a background to exclusive Higgs boson production.
In Fig.\ref{fig:dsig_dMjj_Higgs} we illustrate the situation.
We show both the Higgs signal (hatched area) including experimental
resolution \cite{Pilkington_private,Royon_private} as well as diffractive
$b \bar b$ continuum, QED $b \bar b$ continuum as well as formally
reducible digluon contribution. In the calculation we have assumed
that jet misidentification probability is 1.3\% (which corresponds to the ATLAS misidentification
 factor with a b-tagging efficiency of 60\% \cite{Cox:2007sw}),
i.e. we have multiplied the dijet cross section by a quite small
number 0.013$^2$. The obtained contribution is even larger than the
$b \bar b$ one and overlays the Standard Model Higgs signal. In the
case of exclusive production of Higgs boson beyond the Standard
Model the situation can be better (see e.g.
Refs.~\cite{Cox:2007sw,HKRSTW08,Enberg:2011qh}).

\section{Conclusions}

In the present paper we have discussed the exclusive production of
dijets (both digluon and quark-antiquark ones). We have included the
contribution, previously known from the literature, when both gluons
are emitted from the same $t$-channel gluon line, as well as new
contributions when the gluons are emitted from different $t$-channel
gluon lines (see Fig.~\ref{fig:ggCEP}). We have presented
corresponding formulae with simple prescriptions for the
unintegrated gluon distributions relevant for these two cases. For
both contributions, we made predictions for various differential
distributions at Tevatron (1.96 TeV) and LHC (14 TeV) energies,
including an analysis of different gluon polarisation contributions
and theoretical uncertainties.

The diagram B contribution turned out to be much smaller
than the one (diagram A) known from the literature. They become
comparable only for large rapidity differences of both jets where
the cross section is rather small or when the jet transverse momenta
are small. The latter case can be, therefore, very important for the
central diffractive production of pions. This will be discussed
elsewhere.

We have found that jets corresponding to the mechanism of diagram B
are strongly correlated in their rapidities. This is a consequence
of a specific behavior of off-diagonal gluon distributions in the
ERBL region where $|x|<|\xi| \ll 1$.

We have compared our results with the CDF collaboration data. Our
cross sections are somewhat larger than those obtained in the
literature. Compared to those calculations, we have performed
integration starting from smaller lower limit for screening gluon
transverse momenta and with the choice of scale in the Sudakov form
factor as advocated by Coughlin and Forshaw \cite{CF09}. Our
observation may mean e.g. that the gap survival probabilities are
smaller than usually assumed for this process. Clearly, further work
on this issue is required.

We have discussed also the dijet reducible background to the central
exclusive Higgs boson production. In the framework of the same
model, we have calculated the contribution of the quark-antiquark
jets CEP. We have found that this contribution is much smaller than
that for the gluon-gluon dijets in the whole phase space. However,
the quark contribution, especially the $b \bar b$ one, is very
important, as it constitutes an irreducible background for exclusive
production of the Higgs boson. On the other hand, the gluonic jets
can be misidentified as the $b \bar b$ jets and in this sense they
also contribute to a background for exclusive production of the
Higgs boson.

If both gluonic jets are misidentified as $b$ or anti-$b$ quark
jets, then this leads to an extra background to exclusive Higgs
boson production when the Higgs boson is observed in the $b \bar b$
decay channel. This extra contribution can be even more important
than the irreducible $b \bar b$ contribution. When the realistic
ATLAS misidentification factor is included one obtains the total
background which significantly exceeds the Higgs signal. This
observation suggests that the experimental observation of the
exclusive Standard Model Higgs production may be very challenging.
The situation may be better for beyond the Standard Model Higgs
boson production though, but corresponding detailed analysis
including a Monte-Carlo simulation of backgrounds still needs to be
done.

\vspace{1cm}

{\bf Acknowledgments}

We are particularly indebted to Igor Ivanov for valuable discussions on
theoretical issues related to UGDFs and the Sudakov form factor.
Useful discussions and helpful correspondence with Jean-Rene Cudell,
Rikard Enberg, Krzysztof Golec-Biernat, Gunnar Ingelman, Valery
Khoze, Alan Martin, Andy Pilkington, Christophe Royon, Mikhail
Ryskin, Rafa{\l} Staszewski and Marek Tasevsky are gratefully
acknowledged. This study was partially supported by the Carl Trygger
Foundation and by the polish grants of MNiSW N N202 249235 and N
N202 237040.


\end{document}